\documentclass[epj]{svjour}
\usepackage{graphics}
\usepackage[titletoc]{appendix}
\usepackage{xcolor} 
\usepackage{hyperref}
\usepackage[normalem]{ulem}
\usepackage[square,numbers]{natbib}
\usepackage{amsmath}
\usepackage{amssymb}

\colorlet{mdtRed}{red!50!black}

\newcommand{\dc}[1]{\textsf{\color{red}{ #1}}}
\newcommand{\sg}[1]{\textsf{\color{green}{ #1}}}
\newcommand{\js}[1]{\textsf{\color{blue}{ #1}}}

\begin{document}
%
\title{Imposing multi-physics constraints at different densities on \\
the Neutron Star Equation of State}
\author{Suprovo Ghosh \inst{1}, Debarati Chatterjee \inst{1} \thanks{\emph{debarati@iucaa.in}} 
\and J\"urgen Schaffner-Bielich\inst{2}
}                     
\offprints{}          
\institute{Inter-University Centre for Astronomy and Astrophysics, \\
Post Bag 4, Ganeshkhind, Pune University Campus,\\
Pune - 411007, India
\and 
Institut f\"ur Theoretische Physik, \\
Goethe Universit\"at,\\
Max von Laue Str. 1,
60438 Frankfurt am Main, Germany}
\date{Received: date / Revised version: date}
%
\abstract{
Neutron star matter spans a wide range of densities, from that of nuclei at the surface to exceeding several times normal nuclear matter density in the core. While terrestrial experiments, such as nuclear or heavy-ion collision experiments, provide clues about the behaviour of dense nuclear matter, one must resort to theoretical models of neutron star matter to extrapolate to higher density and finite neutron/proton asymmetry relevant for neutron stars. In this work, we explore the parameter space within the framework of the Relativistic Mean Field model allowed by present uncertainties compatible with state-of-the-art experimental data. We apply a cut-off filter scheme to constrain the parameter space using multi-physics constraints at different density regimes: chiral effective field theory, nuclear and heavy-ion collision data as well as multi-messenger astrophysical observations of neutron stars. Using the results of the study, we investigate possible correlations between nuclear and astrophysical observables.
\PACS{
      {PACS-key}{discribing text of that key}   \and
      {PACS-key}{discribing text of that key}
     } 
} 

\authorrunning {S. Ghosh, D. Chatterjee and J. Schaffner-Bielich}
\titlerunning{Imposing multi-messenger constraints on the Neutron Star EoS}

\maketitle
\section{Motivation}
\label{sec:intro}

Neutron stars (NS) provide us the unique opportunity to study cold and dense matter under extreme conditions far beyond the reach of terrestrial experiments. Having a mass of up to 2 $M_{\odot}$ (solar mass) confined within a radius of roughly 10 km, NSs are among the densest objects in the Universe. As the density increases from the surface towards the interior, one encounters different forms of matter \cite{Lattimer2004,Vidana2018,Lattimer2007,Baym2018,Heiselberg2018}. The NS crust comprises of aggregates of nuclei that become more and more rich in neutrons, until they dissolve in a homogeneous liquid,
with baryonic densities comparable to those reached in heavy-ion collisions at relativistic energies.
At such high densities, strangeness containing exotic particles, such as hyperons and kaons, which are observed only briefly in particle accelerators, can become stable components of NS matter due to weak equilibrium~\cite{Schaffner93,Schaffner96,VidanaEPJA}.   
\\

The behaviour of matter can be represented in terms of its pressure-density relationship, also known as the {\it Equation of State} (EoS) \cite{Baym2018,Heiselberg2018,Lattimer2012}. NSs probe a complementary regime of the phase diagram of QCD (Quantum Chromodynamics, the theory of strong interactions), than those of terrestrial experiments~\cite{Baym2018,QCDBook}. Laboratory nuclear experiments provide information about the cold ($T=0$) EoS close to saturation nuclear density ($n_0 \sim$ 0.16 fm$^{-3}$)~\cite{Lattimer2015,Gandolfi2015}. Heavy Ion Collision (HIC) experiments, such as at GSI in Darmstadt~\cite{FOPI,ASY_EOS}, probe hot and dense but more or less isospin symmetric (in neutron/proton ratio) matter at intermediate densities $\sim 2-3 n_0$ ~\cite{Dexheimer2021,Tsang2018}. NSs on the other hand describe nuclear matter as highly isospin asymmetric (ANM), at high densities and low temperatures. While the above terrestrial experiments provide clues to the behaviour of NS matter around or below saturation density, NS models require extrapolation to higher densities and asymmetries, giving rise to uncertainties in the model parameters.   
\\

Many theoretical schemes are applied to describe the EoS of dense NS matter~\cite{OertelRMP}, such as {\it ab-initio} (where one has to solve the many-body problem)~\cite{MACHLEIDT19871,Haidenbauer2005,Rijken1999,Akmal1998} and {\it phenomenological} (where model parameters are fitted to experimental nuclear observables), including both non-relativistic and relativistic approaches~\cite{Stone2007,Vautherin1972,Serot1986,Serot1997,Dutra2012}. Chiral Effective Field Theory ($\chi$EFT) is a microscopic ab-initio technique which provides an effective and reliable description of pure neutron matter (PNM) at densities $\sim 0.5 - 1.4 n_0$~\cite{Drischler}. In phenomenological NS models such as the Relativistic Mean Field (RMF) model~\cite{Serot1986,Serot1997}, the parameters are fitted to nuclear empirical observables measured at saturation density (e.g. binding energy at saturation, compressibility, symmetry energy and its slope, effective nucleon mass), which are determined from the measurements of neutron skin thickness of $^{208}$Pb or $^{48}$Ca~\cite{Roca2011,Centelles2009,Warda2009}, from electric dipole polarizability $\alpha_D$, from isoscalar giant monopole and dipole resonances~\cite{Shlomo,GARG2007}, nuclear masses~\cite{Atkinson,Kejzlar,Wang2012,Angeli2013} and from isobaric analog states~\cite{DANIELEWICZ2009}. 
\\

On the other hand, constraints for NS models at high densities come from astrophysical observations \cite{Lattimer2012,Lattimer2015,Ozel2016}. NSs are observable throughout the electromagnetic spectrum, from gamma-rays to radio frequencies. With rapid advance in technology, the current generation of space-based and ground-based observatories (e.g. Fermi, INTEGRAL, Parkes pulsar timing array, LOFAR, NICER, XMM-Newton, Chandra) 
are providing us with a vast
pool of Multi-Wavelength (MW) astrophysical data, and a wealth of information about these objects. Many more such MW astronomy projects are under contruction or in conception (Thirty Meter Telescope, E-ELT, SKA, THESEUS, ATHENA, CTA). 
\\

There are many different astrophysical observables concerning NSs that may be inferred using MW astronomy. Spin frequency of NSs can be measured very accurately using pulsar timing. General Relativistic (GR) effects in pulsars in binaries can lead to a very accurate determination of NS masses. It is now well known that NSs can have masses as high as 2 M$_{\odot}$~\cite{Demorest,Antoniadis,Fonseca}.
Recent observations of the heaviest known pulsar PSR J0740+6620 indicate that the maximum mass of the neutron stars should exceed $2.14^{+0.10}_{-0.09}$ $M_{\odot}$ ~\cite{Cromartie}. This value has recently been revised to $2.08^{+0.07}_{-0.07}$ $M_{\odot}$ \cite{fonseca2021refined}. NS radii may be determined using hot spots on NS surface, offset from the rotating pole modulated by stellar rotation \cite{Ozel2010,Steiner2013,Guillot2013,Guillot2014,LattimerSteiner2014}. One of the primary goals of the next generation of hard X ray timing instruments as well as the recently launched NICER (Neutron star Interior Composition ExploreR) mission~\cite{NICER} is to measure NS radii with different masses to high accuracy. 
Recent NICER data provide the mass and radius measurements for pulsars PSR J0030+0451  \cite{NICER0030_Miller,NICER0030_Riley} and PSR J0740+6620 \cite{NICER0740_Miller,NICER0740_Riley}.
Using equations of hydrostatic equilibrium, the Tolman-Oppenheimer-Volkoff equation (TOV), one may map the nuclear EoS to observable quantities such as its mass and radius. Then the measurement of the mass and radius effectively allows us to constrain the NS EoS. 
\\

With the advent of the LIGO-Virgo-Kagra (LVK) global network of gravitational wave detectors~\cite{AdvLIGO2015,Abbott2016,AdvVIRGO2014,Kagra}, a new window to the Universe has opened up, and the pool of data is expected to be further enriched with the upcoming generation of detectors such as the Cosmic Explorer, Einstein telescope or LIGO-India. The most exciting and revolutionary astrophysical discovery in recent times has been the direct detection of Gravitational Waves (GW) from NSs ~\cite{Abbott2017,Abbott2020,Abbott2021}. The detection of the NS merger event GW170817 is particularly unique because this event seen in gravitational waves and its subsequent followup in electromagnetic counterparts by telescopes such as Fermi, Integral and Chandra opened up a new era of Multi-Messenger (MM) astronomy. The role of dense matter EoS is crucial in the tidal response of NSs during late stage of binary inspiral and post-merger remnant oscillations \cite{Hinderer}.
Recent analyses of the GW170817 event~\cite{Abbott2018,De,Abbott2019} and updated values in the recent catalog paper \cite{Abbott_GWTC1} put a tight constraint on the effective tidal deformability $\Lambda$.
Abbott~et~al. (2019)~\cite{Abbott2019} provide confidence intervals of the effective tidal deformability depending on the spin prior and type of interval. For the low-spin case, the highest posterior density interval yields a value of tidal deformability to be $300^{+420}_{-230}$.
As the tidal deformability depends on the stellar mass and radius, and therefore this result also leads to a constraint on the mass-radius relation ~\cite{Abbott2018,Most,Annala}.
\\

Since the pioneering works by Steiner et al. \cite{Steiner2010}, there have been many attempts to impose constraints on the EoS by using multi-messenger observations of neutron stars using a statistical Bayesian scheme~\cite{pang2021,Coughlin2019,Biswas2021,biswas2021prex,biswas2021bayesian,Dietrich2020,O_Boyle_2020}. The idea of this scheme is to match the low density EOS constrained by theoretical and experimental nuclear physics with parametrized high density EOSs satisfying gravitational wave and electromagnetic data ~\cite{Capano2020,Tews_2018,Tews_2019a,Tews_2019b,Gandolfi_2019}. Usually EoSs based on different parametrization schemes such as piecewise polytropes~\cite{Annala,Hebeler2013,Read2009,Gamba2019}, spectral representation~\cite{Fasano2019,Lindblom2018} or speed-of-sound parametrization~\cite{Tews_2018,Greif2019,Landry2020} have been used for such studies, convenient for numerical relativity simulations and parameter estimation from gravitational waveforms. 
Non-parametric inference schemes have also been developed recently~\cite{Landry2019,Legred2021} and combined with chiral EFT results~\cite{Essick2020}. However in such studies the degrees of freedom at intermediate or high densities are overlooked and the EoS curves are modelled by chosen functional forms or considering general physical conditions such as causality. The role of the underlying nuclear physics, such as the key nuclear saturation parameters that control the behaviour of NS global observables, are not apparent from such studies. Few recent studies have used Bayesian scheme within the RMF model \cite{Traversi2020} or hybrid (nuclear + piecewise polytope) parametrizations \cite{Biswas2021} to obtain posterior distributions of empirical parameters but no clear correlation between the nuclear empirical and astrophysical observables was established in such studies.
A few recent investigations employed Bayesian techniques to explore correlations among empirical nuclear parameters and a few NS observables ~\cite{Zhang2019,Xie2019,Carson2019,zimmerman2020measuring}. Some of these works employed nuclear meta-modelling technique \cite{Guven2020}, but such a parametrization is valid around saturation density, and their validity at high densities is questionable.
\\

The aim of this investigation is to apply state-of-the-art data from multiple experimental and astrophysical channels at different densities to constrain the nuclear EoS and to systematically explore possible correlations between astrophysical and nuclear empirical parameters. 
We perform this investigation within the framework of a realistic phenomenological model, the Relativistic Mean Field (RMF) model, to describe NS matter. The uncertainty in the behaviour of nuclear empirical quantities is reflected in the uncertainty in the determination of the RMF models parameters. Motivated by the Bayesian approach, we vary the parameters of the realistic nuclear model within their allowed uncertainties, compatible with state-of-the-art nuclear experimental data. We then apply a simple cut-off filter scheme to constrain the parameter space using a combination of current best-known physical constraints at different density regimes: theoretical (chiral effective field theory) at low densities, experimental (nuclear and heavy-ion collision) at intermediate densities and multi-messenger (multi-wavelength electromagnetic as well as GW) astrophysical data at high densities to restrict the parameter space of the nuclear model. A similar scheme was applied by Hornick~et~al.~\cite{Hornick}, where limits were imposed on only isovector saturation nuclear parameters within the RMF parameter space using chiral effective field theory at low densities and multi-messenger (multi-wavelength electromagnetic as well as GW) astrophysical data at high densities. We extend this study by considering variation of all the parameters within their uncertainties, apply updated multi-messenger constraints and also impose additional constraints from heavy-ion data at intermediate densities.
\\

The structure of the article is as follows: in Sec.~\ref{sec:formalism}, we describe the formalism, particularly the microscopic description (details of the nuclear model) and the macroscopic global modelling. The details of the filter scheme are discussed in Sec.~\ref{sec:bayesian}. We first test the scheme by reproducing the results of \cite{Hornick} in Sec~\ref{sec:tests}. We then extend the scheme to a full analysis including a wider range of parameters and additional constraints in Sec.~\ref{sec:full}. We demonstrate the results of the investigation, including comparison with previous results and the study of correlations. Finally, in Sec.~\ref{sec:discussions} we discuss the main implications of this work.

\section{Formalism}
\label{sec:formalism}

\subsection{Microscopic description}
\label{sec:micro}

As mentioned in Sec.~\ref{sec:intro}, we perform this study within the framework of the RMF model to calculate the NS EoS. We start with the interaction Lagrangian density~\cite{Hornick,Chen,Fattoyev2010}:
\begin{eqnarray}
{\cal{L}}_{int} &=& \sum_N \bar{\Psi_N} \left[ g_{\sigma} \sigma - g_{\omega} \gamma^{\mu} \omega_{\mu} - \frac{g_{\rho}}{2} \gamma^{\mu} \vec{\tau} \vec{\rho_{\mu}} \right] \Psi_N \nonumber \\
&-& \frac{1}{3} b m (g_{\sigma} \sigma)^3 -\frac{1}{4} c  (g_{\sigma} \sigma)^4 \nonumber \\
&+& \Lambda_{\omega} (g_{\rho}^2 \vec{\rho_{\mu}} \vec{\rho^{\mu}} ) (g_{\omega}^2 \omega_{\mu} \omega^{\mu} ) + \frac{\zeta}{4!} (g_{\omega}^2 \omega_{\mu} \omega^{\mu} )^2
\label{eq:lagr}
\end{eqnarray}
where $\Psi_N$ is the Dirac spinor for nucleons $N$, $m$ is the vacuum nucleon mass, while $\gamma^{\mu}$ and $\vec{\tau}$ are the Dirac and Pauli matrices respectively. The interaction among the nucleons is mediated by the exchange of the scalar ($\sigma$), vector ($\omega$) and the isovector ($\rho$) mesons. The isoscalar nucleon-nucleon couplings $g_{\sigma}$ and $g_{\omega}$ are determined by fixing them to nuclear saturation properties. The $\sigma$ meson self-interaction terms $b$ and $c$ ensure the correct description of nuclear matter at saturation.  The~effective nucleon mass is then defined as 
$m^* = m - g_{\sigma} \sigma$. The isovector and mixed $\omega-\rho$ couplings $g_{\rho}$ and $\Lambda_{\omega}$ can be related to empirical quantities such as symmetry energy ($E_{sym}$) and its slope ($L_{sym}$) \cite{Chen}. The quartic $\omega$ self-coupling $\zeta$ is set to zero in this study, because the corresponding term is know to soften the EoS which is in tension with the $2M_\odot$ mass constraint from pulsar data. 
The choice of model parameters is discussed in Sec.~\ref{sec:para}.
\\

From the Lagrangian density Eq.~(\ref{eq:lagr}), one can solve the equations of motion of the constituent particles as well as those of the mesons \cite{Hornick}. In the mean-field approach, the meson fields are replaced by their mean-field expectation values. One can then calculate the EoS starting with this RMF model. The energy density is given by \cite{Hornick}
\begin{eqnarray}
\varepsilon &=& \sum_N \frac{1}{8 \pi^2} \left[ k_{F_N} E_{F_N}^3 +  k_{F_N}^3 E_{F_N} - {m^*}^4 \ln \frac{k_{F_N}+E_{F_N}}{m^*} \right]  \nonumber \\
&+& \frac{1}{2} m_{\sigma}^2 \bar{\sigma}^2 + \frac{1}{2} m_{\omega}^2 \bar{\omega}^2 + \frac{1}{2} m_{\rho}^2 \bar{\rho}^2 \nonumber \\
&+& \frac{1}{3} b m (g_{\sigma} \bar{\sigma})^3 +\frac{1}{4} c  (g_{\sigma} \bar{\sigma})^4 \nonumber \\
&+& 3 \Lambda_{\omega} (g_{\rho} g_{\omega} \bar{\rho} \bar{\omega})^2 + \frac{\zeta}{8} (g_{\omega} \bar{\omega})^4~.
\end{eqnarray}
where $k_{F_N}$ and $E_{F_N}$ are the Fermi momentum and energy of the corresponding nucleon $N$ respectively. The pressure $P$ can be derived from the energy density using the Euler relation 
\begin{equation}
P = \sum_N \mu_N n_N - \varepsilon~,
\end{equation}
where $n$ is the number density and the nucleon chemical potentials ($\mu$) are given by
\begin{equation}
\mu_N = E_{F_N} + g_{\omega} \bar{\omega} 
+ \frac{g_{\rho}}{2} \tau_{3N} \bar{\rho}~.
\end{equation}

\subsection{Macroscopic description}
\label{sec:macro}


Given an EoS, the equilibrium configurations of non-rotating relativistic NSs can be obtained by solving the Tolman-Oppenheimer-Volkoff (TOV) equations of hydrostatic equilibrium \cite{GlendenningBook,schaffner2020},
\begin{eqnarray}
\frac{dm(r)}{dr} &=& 4 \pi \varepsilon(r) r^2 ~, \nonumber \\
\frac{dp(r)}{dr} &=& - \frac{[p(r) + \varepsilon(r)] [m(r)+4 \pi r^3 p(r)] }{r(r-2 m(r))}~, \nonumber \\
\label{eq:tov}
\end{eqnarray}
%
Integrating the TOV equations from the centre of the star to the surface, one can obtain global NS observables, such as mass, $M$, radius, $R$, and compactness, $C=M/R$ that can be deduced from astrophysical data. The boundary conditions that must be satisfied are a vanishing mass, $m|_{r=0}=0$, at the centre of the star, and~a vanishing pressure, $p|_{r=R}=0$), at the surface.
 The tidal deformability, $\Lambda$, can be obtained by solving a set of differential equations coupled with the TOV equations~\cite{Hinderer,Hinderer2008,Damour2009,YagiYunes2013b}. 
It was shown in these works that the tidal deformability parameter $\Lambda$ depends on the mass and radius of the NS (and hence its EoS)~\cite{Hinderer2010}
\begin{equation}
    \Lambda = \frac{2}{3} k_2 \left( \frac{R}{M} \right)^5~,
\label{eq:tidal}
\end{equation}
where $k_2$ is the $l=2$ dimensionless tidal Love number.

\subsection{Model parameters}
\label{sec:para}

To obtain the EoS, one must know the coupling constants, which are referred to as the model parameters. The nucleon isoscalar coupling constants ($g_{\sigma}, g_{\omega}, b, c$) defined in Sec.~\ref{sec:micro} are calculated by fixing the empirical properties obtained from nuclear experiments, such as nuclear saturation  density ($n_0$), binding energy per nucleon ($E/A$), incompressibility  ($K_{sat}$) and the effective nucleon mass  ($m^*$) at saturation. On the other hand, isovector coupling constants  ($g_{\rho_N},\Lambda_{\omega}$) are obtained by fixing the symmetry energy  ($E_{sym}$) and the slope  ($L_{sym}$) of symmetry energy at saturation  (see e.g., \cite{Chen,Hornick}).
\\

In order to test our calculations, we first calibrated the model parameters by calculating the EoS, corresponding mass-radius relations and tidal deformability for the well-known parametrizations GM1 and GM3 \cite{GM}. We then reproduced the EoS for neutron matter and the mass-radius relations given in Hornick et al.~\cite{Hornick} (see e.g.~Figs.~1, 2, 5, 6 of that paper). The saturation properties of these parameter sets are recalled in Table~\ref{tab:emppara}. 
\\

\begin{table*}
\caption{Empirical saturation parameters for the RMF models GM1, GM3
\cite{GM} and the parameter sets from HTZCS \cite{Hornick}.}
\label{tab:emppara}       
\begin{center}
\begin{tabular}{lllllll}
\hline\noalign{\smallskip}
   Model & $n_0$ & $E_{sat}$ & $K_{sat}$ & $E_{sym}$ & $L_{sym}$ & $m^*/m$ \\
    {} & ($fm^{-3}$) & (MeV) & (MeV) & (MeV) & (MeV) & {}\\
\noalign{\smallskip}\hline\noalign{\smallskip}
\hline
{GM1} & 0.153 & -16.3 & 300 & 32.5 & 93.7 & 0.70 \\
\hline
{GM3} & 0.153 & -16.3 & 240 & 32.5 & 89.7 & 0.78 \\
\hline
{HTZCS} & 0.150 & -16 & 240 & {30-32} & {40-60} & 0.55-0.75 \\
\noalign{\smallskip}\hline
\end{tabular}
\end{center}
\vspace*{1cm}  
\end{table*}

\section{Cut-off filter scheme}
\label{sec:bayesian}

In a typical Bayesian analysis, an initial range of parameters is varied to produce a {\it prior} probability distribution. The {\it posterior} distribution is then obtained by multiplying the prior and likelihood functions. The likelihood functions are appropriately chosen physical conditions.
\\

In this study, we apply a ``cut-off filter" scheme, where we impose strict limits on the nuclear parameter space using multi-density constraints. The nuclear empirical parameters are varied randomly within their uncertainty range in Table~\ref{tab:emppara} to generate the EoSs. Among the EoSs, we then allow only the ones which satisfy certain physical requirements or ``filters", described in detail in Sec.~\ref{sec:filters}.
\\

Although Miller~et~al. (2019)~\cite{2019Miller} point out the statistical uncertainties in constraining EoS by putting strict limits from multi-messenger observations of neutron stars like we have used here, using the cut-off scheme gives a correct estimate of the nuclear parameter ranges consistent with the observations. Many recent works~\cite{Annala2017llu,Most2018hfd,2020Annala,annala2021multimessenger,refId0} also used similar cut-off schemes for constraints for ultra-dense matter. Further, one must note that chiral effective theory and heavy-ion collision experiments give a band of nuclear observables as constraints and a statistical weighting of the model-dependent heavy-ion data might lead to a false confidence in the results. So, using a cut-off filter scheme for the constraints seems appropriate in these cases, in order to extract the underlying physics. We have also generated a very large number of prior EoSs to constrain the parameter space. One may verify whether Bayesian analysis with statistical weighting affects the posterior probability distribution, or significantly alters the physical correlations between nuclear empirical parameters and astrophysical observables which is the main aim of this study. In Sec.~\ref{sec:weight} and Sec.~\ref{sec:weight_hic} we check this explicitly by redoing the analysis with a proper calculation of the likelihood of the data.
\\

As mentioned in the introduction, such limits were also imposed on the RMF parameter space in the recent study by Hornick~et~al.~\cite{Hornick}. A similar scheme using the meta-modelling approach~\cite{Margueron2018a} was also employed in the analysis of the effect of uncertainties in nuclear empirical parameters and the study of their correlations with NS observables~\cite{Margueron2018b} and experimental observables of finite nuclei by one of the authors (D.C.) of this work~\cite{Chatterjee2017}. This work~\cite{Chatterjee2017} found no correlation between the symmetry energy $E_{sym}$ and its first order derivative $L_{sym}$, which is at variance with several studies in the literature which find a good correlation between $E_{sym}$ and $L_{sym}$ \cite{Tsang,Kortelainen,Ducoin11}.

\subsection{Filter functions}
\label{sec:filters}

The following physical constraints from multi-disciplinary physics are applied in this work at different densities, as described below:

\begin{itemize}
    \item \textbf{At low densities: $\chi$EFT} \\
   Important constraints on the EoS of NS matter at low baryon densities $n_b$ in the range of $\sim 0.5-1.4 n_0 $ come from microscopic calculations of pure neutron matter (PNM) using chiral effective field theory \cite{Drischler,Hebeler2013,Carbone2013}. Chiral EFT applies low momentum expansion of nuclear forces related to the symmetries of QCD. It accounts for many-body interactions among nucleons using order-by-order expansions in terms of contact interactions and long-range pion exchange interactions. Thus the EoS of PNM can be compared with the latest $\chi$EFT calculation results~\cite{Tews_2018}.\\
   
    \item \textbf{At high densities: astrophysical observations}\\ 
    The constraints at high density come from multi-messenger astrophysical observations, such as high mass NS observations and
    GW constraints on tidal deformability, as follows : 
    
\begin{enumerate}
    \item From the recent observations of the heaviest known pulsar PSR J0740+6620, the maximum mass of the neutron stars should exceed $2.08^{+0.07}_{-0.07}$ M$_{\odot}$~ \cite{fonseca2021refined}. This sets an upper bound on the maximum NS masses corresponding to the EoSs considered in this study.
     \item 
     Using the low-spin highest posterior density interval for tidal deformability from the recent analyses of the GW170817 event~\cite{Abbott2019}, we apply a constraint on the upper bound of the effective tidal deformability $\Lambda <$ 720~\cite{Tong2020}. We do not consider the lower limit on tidal deformability in this study. 
As explained in Sec.~\ref{sec:intro}, the tidal deformability depends on the mass and radius (see Eq.~\ref{eq:tidal}), and therefore this result also leads to a constraint on the mass-radius relation, imposing that NSs should have $R_{1.4M_{\odot}} < 13.5$ km~\cite{Most,Annala}.
\\
\\

\end{enumerate}

\item \textbf{At intermediate densities: heavy-ion collisions} \\
  At intermediate densities in the range $1-3 n_0$, we have additional information from heavy-ion collision experiments.
  There are several heavy-ion collision experiments, which impose important constraints on the nuclear parameters in the density region of $n_b/n_0 \sim 1-3$ which we discuss in detail below:\\
  
\begin{enumerate}
\item \textbf{KaoS experiment}
        The measurements on $K^+$ meson production in nuclear
collisions at subthreshold energies were performed with
the Kaon Spectrometer (KaoS) at GSI, Darmstadt~\cite{Hartnack}. The Kaon multiplicity in Au+Au \& C+C collisions is an indicator of the compressibility of dense matter at $n_b \sim 2-3 n_0$. The level of compression is in turn controlled by the stiffness of nuclear matter through the nucleon potential $U_N$. The more attractive $U_N$ is, the higher is the produced $K^+$ meson abundance, which therefore can serve as a probe for the stiffness of nuclear matter. The analysis of the experimental results using IQMD transport models indicated that the EoS in this density range must be soft, as described by a simple Skyrme ansatz with an incompressibility $K$ of $\sim$ 200 MeV or less \cite{Hartnack,Fuchs2001}. \\

\item \textbf{FOPI experiment}
        Using the data on elliptic flow in Au+Au collisions between 0.4 and 1.5A GeV by FOPI collaboration \cite{FOPI}, one can obtain constraints for the EoS of compressed symmetric nuclear matter (SNM) using the transport code Isospin Quantum Molecular Dynamics (IQMD) by introducing an observable describing the evolution of the size of the elliptic flow as a function of rapidity. This observable is sensitive to the nuclear EoS and a robust tool to constrain the incompressibility of nuclear matter up to 2~$n_0$. Using the FOPI data, one can obtain a constraint for the binding energy of SNM in the density region of $n_b/n_0 \sim 1.4-2.0$ ~\cite{FOPI}. 
        \item \textbf{ASY-EOS experiment}
        Directed and elliptic flows of neutrons and light charged particles were measured for the reaction $^{197}$Au+$^{197}$Au at
        400 MeV/\\ nucleon incident energy within the ASY-EOS experimental campaign at the GSI laboratory, Germany \cite{ASY_EOS}. From the comparison of the elliptic flow ratio of neutrons with respect to charged particles with UrQMD predictions, a value $\gamma = 0.72 \pm 0.19 $ is obtained for the power-law coefficient describing the density dependence of the potential part in the parametrization of the symmetry energy, as defined in the equation below~\cite{ASY_EOS} :
\begin{eqnarray}
&& E_{sym} = E_{sym}^{pot} + E_{sym}^{kin} \nonumber \\
&=&  E_{sym}^{pot}(n_0) (n_b/n_0)^{\gamma} + 12 \textrm{MeV} (n_b/n_0)^{2/3}~.
\end{eqnarray}
It represents a new and more stringent constraint on the symmetry energy for ANM in the regime of supra-saturation density. The densities probed are shown to reach beyond 2 $n_0$. 
\\

We would like to emphasize here that although the constraints from heavy-ion collisions are important and interesting, they are model-dependent. The heavy-ion data are confronted by well established transport codes such as IQMD~\cite{AICHELIN1991,Hartnack1998} or Ultra relativistic Quantum Molecular Dynamics (UrQMD) using various phenomenological EoSs to assess which EoS is most compatible with the data. There are two important points to be aware of in this regard. First of all, the data that are claimed to constrain the EoS of nuclear matter require confirmation by independent experimental efforts. Secondly, it must be proved that the conclusions using a particular transport code (IQMD or UrQMD) are not limited to that particular code. Some efforts have already been made in this direction~\cite{Hartnack2012}. In view of the points raised earlier we do not consider the constraints from collective flow of nucleons~\cite{Danielewicz}. 
    \end{enumerate}
\end{itemize}

\subsection{Correlations}
\label{sec:corr}

Using the filtered EoSs obtained from the analysis, one can then perform a search for physical correlations of the different empirical parameters among themselves, as well as with the NS astrophysical observables such as radii and tidal deformability of canonical 1.4~M$_{\odot}$ and massive 2~M$_{\odot}$ NSs. For this study, we use Pearson's linear correlation coefficient defined as : 
\begin{equation}\label{corrcoff}
    R_{XY} = \frac{Cov(X,Y)}{S(X)S(Y)}
\end{equation}
between two random variables $X$ and $Y$ where $Cov(X,Y)$ denotes the covariance and $S(X),S(Y)$ denote the standard deviation for the variables $X$ and $Y$ respectively. 
The correlations studied here are calculated from the confidence bounds used in this study.

\subsection{Statistical weighting}
\label{sec:weightcorr}
In Sec.~\ref{sec:bayesian}, we discussed the need to take into account the statistical uncertainties in the measurement of the nuclear and astrophysical observables while constraining the EoS. In this section, we summarise how to include the effects of statistical weighting on the correlations.
\\
In this work, we use a $\chi^2$ distribution, as is done for hypotheses testing or goodness of fit~\cite{chi2_Chen,Chatterjee2017,chi2_Reinhard}. We obtain our priors from the parameter space \{P\} by uniformly varying the nuclear parameters from Table~\ref{tab:parafull}. The posterior is obtained by multiplying the prior and the likelihood functions. For Gaussian variables, the likelihood is proportional to
\begin{equation}\label{weight}
    W = e^{-\chi^2/2}
\end{equation}
which we assign as weight $W$ to each parameter set \{P\}. In ~\eqref{weight}, the $\chi^2$ is a weighted sum of squared deviations defined as below :  
\begin{equation}\label{chi2}
    \chi^2 = \sum_i \frac{(O_i - C_i)^2}{\sigma_i^2}
\end{equation}
As discussed in Sec.~\ref{sec:filters}, for low density $\chi$EFT filters, we get bounds for the PNM EoS in the density range of $\sim 0.5-1.4 n_0$~\cite{Drischler}. We use these bounds as the uncertainty range to calculate the standard deviation($\sigma_i$). For astrophysical observations, mass measurement of PSR J0740+6620~\cite{fonseca2021refined} ($2.08^{+0.07}_{-0.07}$ M$_{\odot}$) and tidal deformability measurement from GW170817~\cite{Abbott2019}($300^{+420}_{-230}$) give the observed mean($O_i$) and $\sigma_i$ values for the observables. Using the equations~\eqref{chi2} and \eqref{weight} we then calculate the weights associated with each parameter set \{P\} for these filters. For $\chi EFT$ and multi-messenger observations of the NSs, the uncertainties can be determined to a good accuracy, but due to the model dependence of the heavy-ion collision data, statistical weighting may lead to false conclusions. We include weights for HIC filters also in Sec.~\ref{sec:weight_hic} with a word of caution. As discussed in Sec.~\ref{sec:hic}, for KaOS constraint, we calculate the nucleon potential and associated errors($\sigma_i$) from the non-relativistic Skyrme model for $K_{sat}$=200 MeV and $n_0 = 0.15 fm^{-3}$ and use it as a upper limit~\cite{Sagert} to calculate the weight. From FOPI~\cite{FOPI} and ASY-EOS~\cite{ASY_EOS} experiments, we obtain uncertainty bands for binding energy and Symmetry energy respectively which are used to calculate the associated mean and standard deviations in the corresponding density range. The final weight for a parameter set \{P\} is the product of the weights associated with each constraint: 
\begin{equation}
    W_{\{P\}} = W_{\chi EFT} \times W_{Astro} \times W_{HIC}
\end{equation}

The definition of the Pearson's linear correlation coefficient in Eqn.~\eqref{corrcoff} also gets modified when we include weights. The definition of covariance between two variables X and Y with a weight W is 
\begin{eqnarray}\label{Covweight}
  &&  Cov(X,Y; W)  \nonumber \\
  &=& \frac{\sum_i W_i  (X_i - M(X;W)) (X_i - M(X;W))}{\sum_i W_i}
\end{eqnarray}
where $M(X,W)$ is the mean of the variable X defined as $M(X,W) = \frac{\sum_i X_iW_i}{\sum_i W_i}$. The weighted correlation coefficient is defined as 
\begin{equation}\label{weightcorrcoff}
    R_{XY;W} = \frac{Cov(X,Y; W)}{\sqrt{Cov(X,X; W)}\sqrt{Cov(Y,Y; W)}}
\end{equation}

\section{Preliminary results: Testing the scheme}
\label{sec:tests}


In order to test whether the applied scheme can be used to impose constraints on the EoS parameter space, we first reproduce the results from the parameter study in Hornick et al.~\cite{Hornick}, hereafter referred to as HTZCS. In this work, the isoscalar couplings ($n_0, E_{sat}, K_{sat}$), which have comparatively lower uncertainties, were kept fixed while the isovector couplings ($E_{sym}, L_{sym}$) along with effective mass $m^*/m$ were varied individually within the chosen range motivated by theoretical and experimental uncertainties (see Table~\ref{tab:emppara}), i.e. to obtain the EoSs.
\\

\subsection{Input EoSs}
\label{sec:prior}
\begin{figure}[htbp]
\resizebox{0.5\textwidth}{!}{\rotatebox{0}%
{\includegraphics{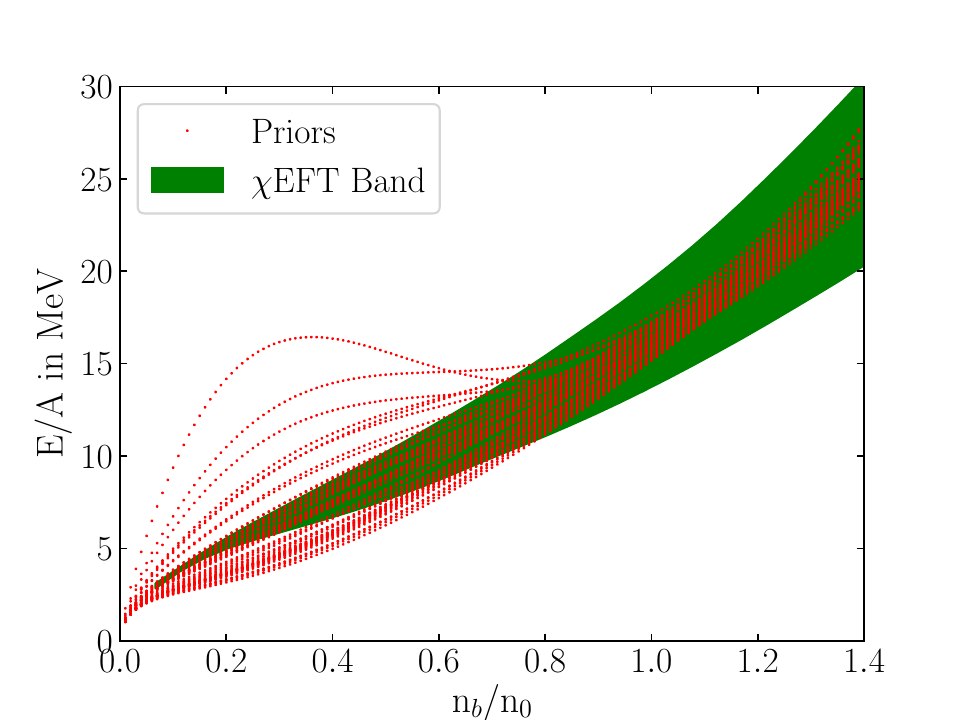}}
}
\caption{Binding energy $E/A$ of pure neutron matter EoS as a function of normalized baryon density $n_b/n_0$ for the prior distribution, superimposed with the chiral effective field theory ($\chi$EFT) band}~\cite{Drischler}.
\label{fig:testprior_alldens_binden}     
\end{figure}

We initially obtain input EoSs by randomly picking up values within the parameter space from HTZCS given in Table~\ref{tab:emppara}. For these sets, we calculate the PNM and ANM EoS as described previously. In Fig.~\ref{fig:testprior_alldens_binden}, we plot the binding energies vs density of low density PNM EoS superposed with the $\chi$EFT band~\cite{Drischler}.


\subsection{Output EoSs}
\label{sec:posterior}

After generating the random EoSs, we use the $\chi$EFT and astrophysical filters described in Sec.~\ref{sec:filters} to obtain filtered sets for the nuclear parameters and NS observables. Around 50\% of the EoSs pass through both the filters (around 60\% for only $\chi$EFT and 80\% for only NS observables.)

\subsubsection{Low density: $\chi$EFT}
\label{sec:CEFT}
\begin{figure}[h]
  \begin{center}
      \resizebox{0.5\textwidth}{!}{%
{\includegraphics{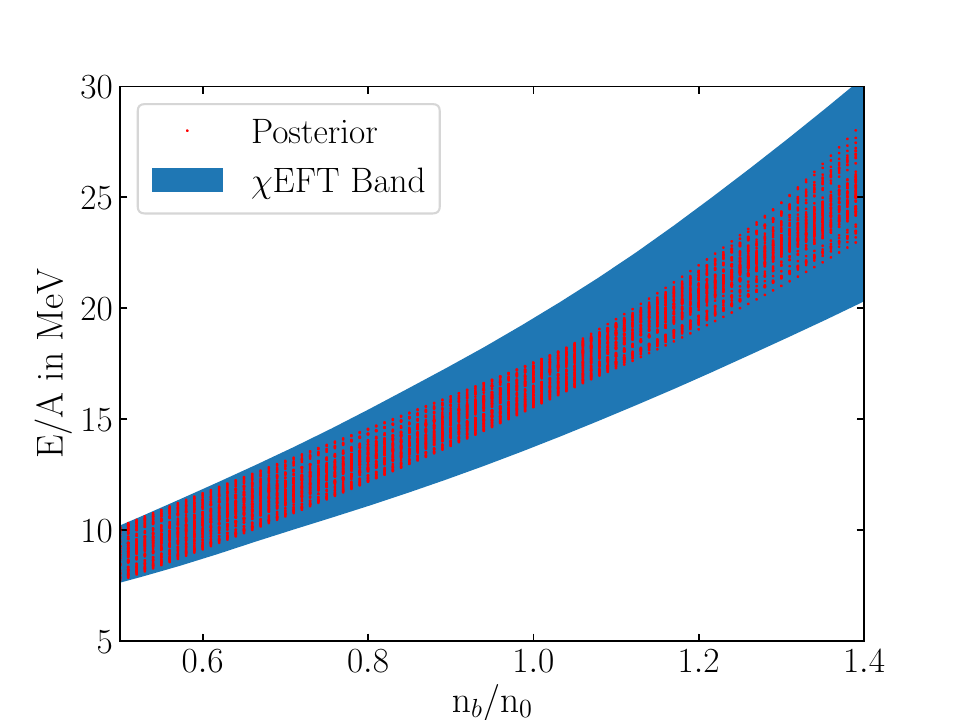}}
}
      \caption{Binding energy $E/A$ of posterior pure neutron matter EoSs as a function of normalized density $n_b/n_0$ allowed by chiral effective field theory ($\chi$EFT) data} \cite{Drischler}.
    \label{fig:testposterior_binden} 
  \end{center}
\end{figure}

We first apply the cut-off filter applying the constraints from $\chi$EFT on the EoS of pure neutron matter~\cite{Drischler}. In order to do so, we evaluate the binding energies in the density range of $n_b/n_0  \sim 0.5-1.4 $ corresponding to the $\chi$EFT data and allow only those parameter sets that lie within the band allowed by $\chi$EFT  calculations (Fig.~\ref{fig:testposterior_binden}). 

\subsubsection{High density: NS astrophysical observations}
\label{sec:filterobs}
  Comparing with the recent observations of mass, radius, moment of inertia and tidal deformability using multi-messenger astrophysical and GW data, we can rule out further combinations of parameter sets and allow only those combinations which simultaneously satisfy all constraints on NS observables. \\
  
\begin{figure}[htbp]
  \begin{center}
 \resizebox{0.5\textwidth}{!}{%
      \includegraphics{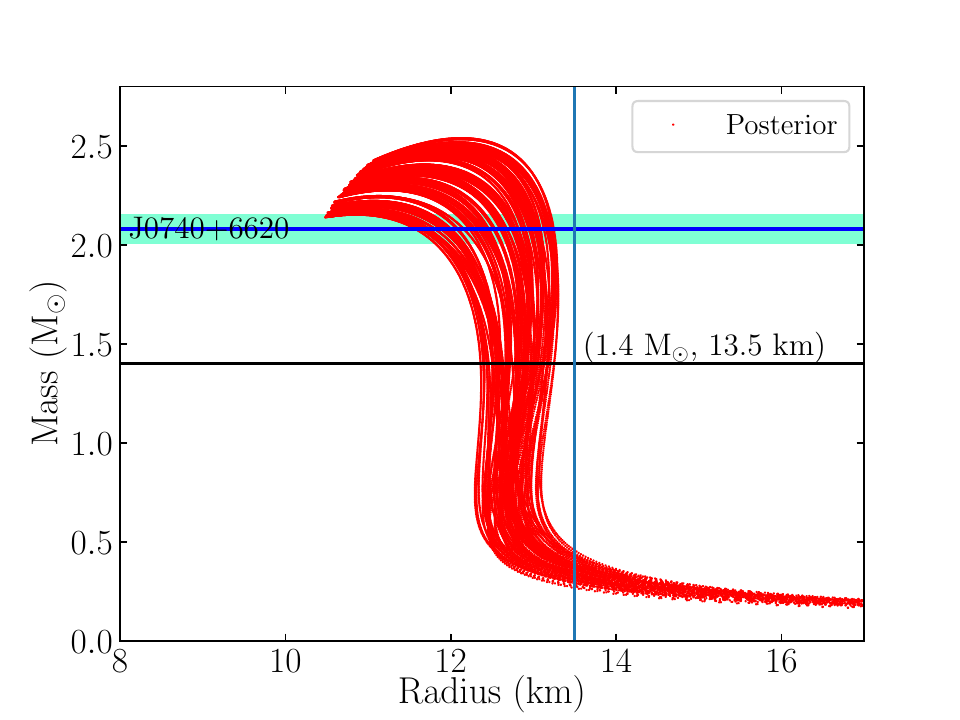}
}
\caption{Mass-radius relations for posterior ANM EoSs after passing through $\chi$EFT and NS observations filters(maximum mass of  PSR J0740+6620~\cite{fonseca2021refined} and tidal deformability $\Lambda$ from  GW170817~\cite{Abbott2019}). The light green band indicates the uncertainty uncertainties in the measurement of maximum mass of PSR J0740+6620~\cite{fonseca2021refined}}
    \label{fig:mr_allsets}
  \end{center}
\end{figure}

\begin{figure}[htbp]
  \begin{center}
  \resizebox{0.5\textwidth}{!}{%
      \includegraphics{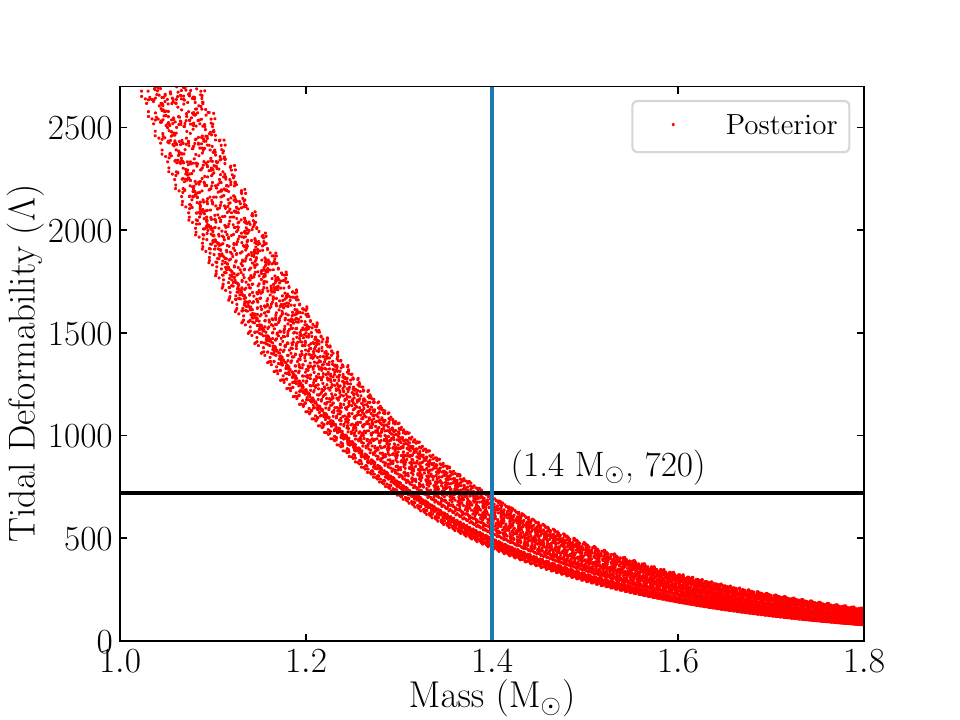}
      }
    \caption{Dimensionless tidal deformability for posterior ANM EoSs after passing through $\chi$EFT and NS observations filters (maximum mass of  PSR J0740+6620~\cite{fonseca2021refined} and tidal deformability $\Lambda$ from  GW170817~\cite{Abbott2019}).
    }
    \label{fig:tidal_allsets}
  \end{center}
\end{figure}

 In Fig.~\ref{fig:mr_allsets}, we plot the mass-radius relations corresponding to the filtered ANM EoSs, along with the uncertainty in the measurement of maximum mass of the most massive pulsar yet observed PSR J0740+6620~\cite{fonseca2021refined}. It is evident from this figure that the NS radii span a wide range from 11-14 km. 
We plot the dimensionless tidal deformability as a function of NS mass in Fig~\ref{fig:tidal_allsets} corresponding to the posterior ANM EoSs.
\\

\subsection{Correlations among nuclear empirical parameters and with NS observables}
\label{sec:corrprimary}

\begin{figure}[htbp]
 \begin{center}
 \resizebox{0.55\textwidth}{!}{%
      \includegraphics{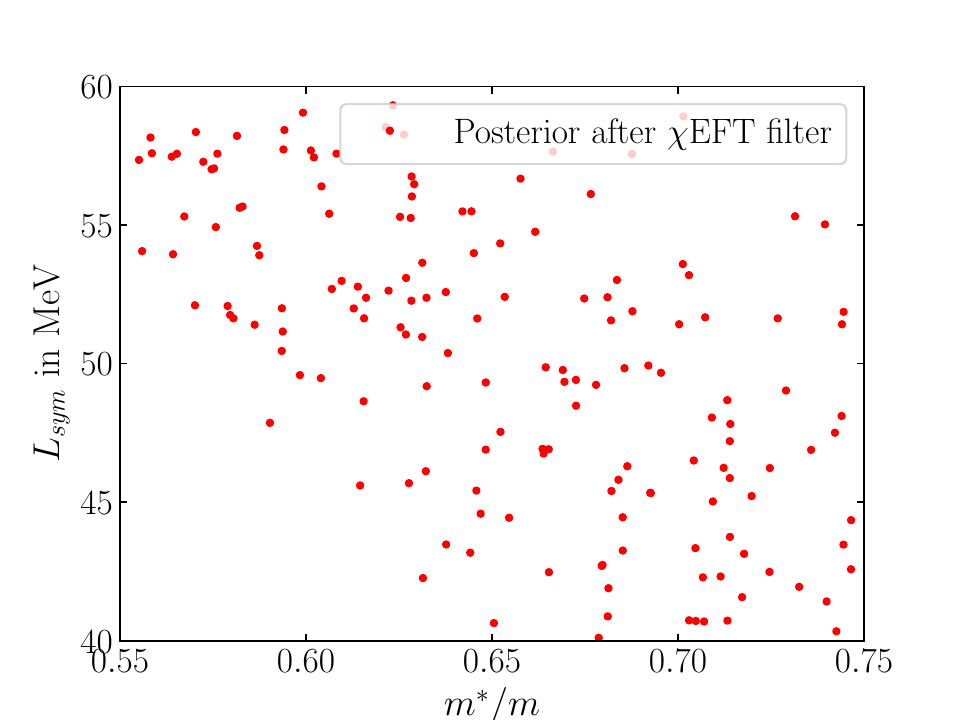}
      }
    \caption{Correlation between $L_{sym}$ and $m^*/m$ from the posterior obtained after the $\chi$EFT  filter}
    \label{fig:corrLm}
  \end{center}
\end{figure}

To verify some of the observations in HTZCS, we probe whether there exist correlations among nuclear empirical parameters after applying the $\chi$EFT band as a filter in our  analysis. 
We plot the correlations among the parameters $L_{sym}$ and $m^*/m$ in 
Fig.~\ref{fig:corrLm}. From the plot, we observe that there are no points in the region below $L_{sym}$ = 55 MeV and $m^*/m$ = 0.65. 
This was also concluded from Fig. 3 and 4 in HTZCS: that it is more difficult to obtain physical solutions for the neutron matter EoS simultaneously for small $L_{sym}$ and $m^*/m$ compatible with the $\chi EFT$ results. This can be explained by the Hugenholtz-van Hove theorem, which
states that the binding energy per particle has to be equal
to the Fermi energy at saturation. The decrease in $m^*/m$ (increase in the scalar potential) would lead to larger values for the vector field and a stiffer EoS. The softening of the EoS for small $L_{sym}$ competes with the stiffening of the EoS as $m^*/m$ is decreased, leading to either a solution outside the band of $\chi$EFT or the appearance of unstable solutions at sub-saturation density.
\\

\begin{figure}[h]
  \begin{center}
  \resizebox{0.55\textwidth}{!}{%
      \includegraphics{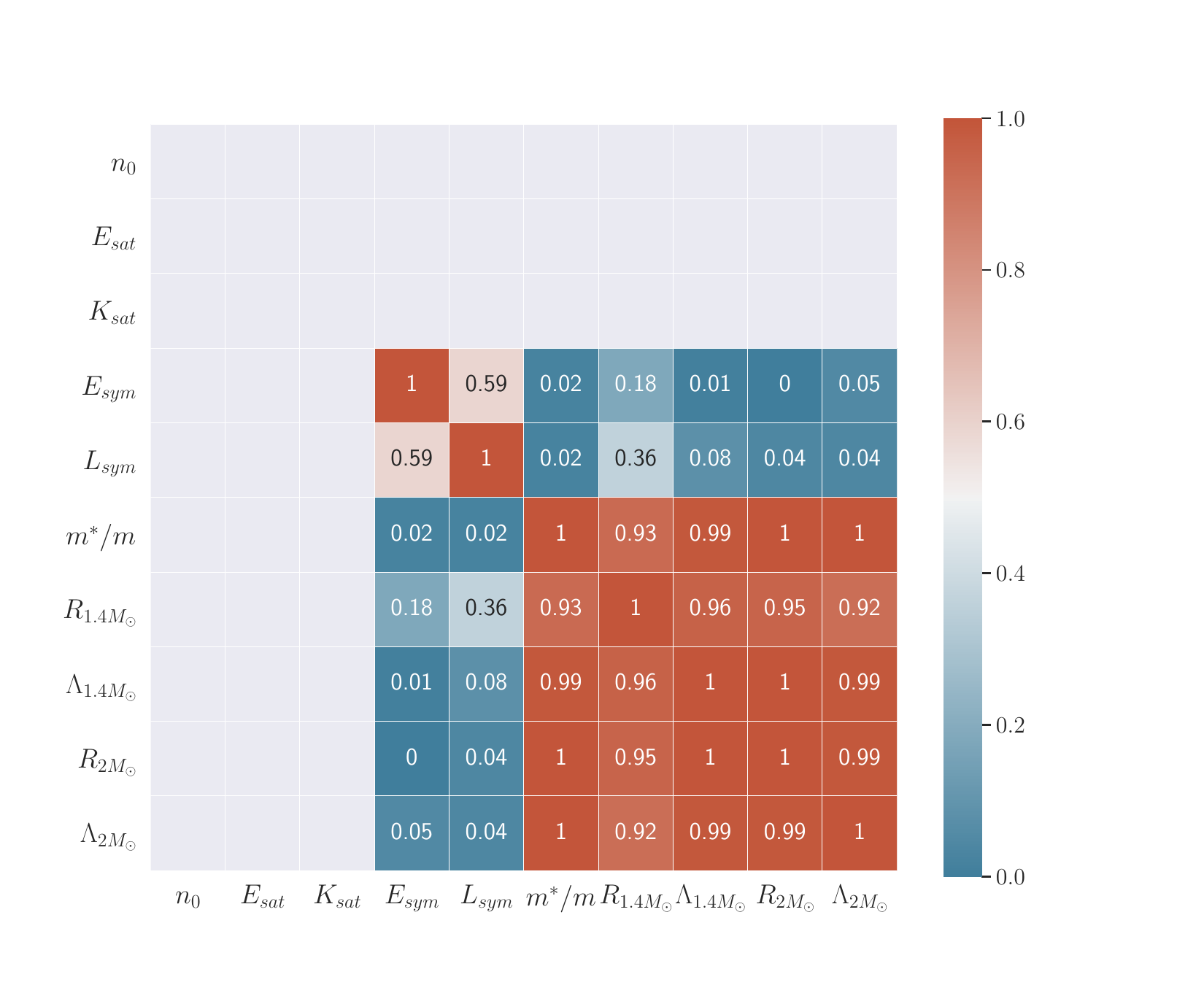}
      }
    \caption{Posterior correlation matrix for variation of isoscalar empirical parameters and NS observables, after application of the $\chi$EFT and NS observations filter (maximum mass of  PSR J0740+6620~\cite{fonseca2021refined} and tidal deformability $\Lambda$ from  GW170817~\cite{Abbott2019})}
    \label{fig:corrfigprimary}
  \end{center}
\end{figure}

 In Fig.~\ref{fig:corrfigprimary}, we display the correlation matrix of the following quantities: nuclear empirical parameters ($E_{sym}$, $L_{sym}$, $m^*/m$) and the NS observables ($R_{1.4M_{\odot}}$, $\Lambda_{1.4M_{\odot}}$, $R_{2M_{\odot}}$, $\Lambda_{2M_{\odot}}$). Some of the crucial observations we make from the correlation matrix are listed below: 

\begin{itemize}
    \item As given in Table.~\ref{tab:emppara}, the values of the isoscalar parameters $n_0$, $E_{sat}$ and $K_{sat}$ are not varied as in HZTCS  (this will be done in the full analysis in Sec.~\ref{sec:full}) and therefore the correlations among them do not appear in the correlation matrix.
     \item  We see a very high correlation between the tidal deformability $\Lambda$ and radius $R$ which is expected from the formula of tidal deformability parameter ($\Lambda \propto R^{5}$) (see Eq.~\ref{eq:tidal}).
        \item The effective nucleon mass is highly correlated with the radius and tidal deformability. As explained in HTZCS, due to the Hugenholtz-van-Hove theorem, the smaller the effective nucleon mass is, the stiffer the EoS is. Thus the effective mass controls both the variation in the maximum mass and the radius, and consequently also the tidal deformability.
      \item Among the nuclear empirical parameters, the symmetry energy $E_{sym}$ and its first order derivative $L_{sym}$ appear to be have a moderate correlation (0.59). 
       \item There is a weak correlation (0.36) between the slope of symmetry energy parameter $L_{sym}$ and radius at 1.4M$_{\odot}$. A correlation between $L_{sym}$ and $R_{1.4 M_{\odot}}$ has been reported in several articles in the literature \cite{Fattoyev,Alam,Zhu,Lim}, although in HTZCS $R_{1.4 M_{\odot}}$ was found to be nearly independent of $L_{sym}$. It was also concluded in HTZCS, that the variation of $m^*/m$ is more important than $L_{sym}$ in the determination of $R_{1.4 M_{\odot}}$. This is clearly in agreement with our results. 
       \item We also note that the correlation of $L_{sym}$ with $R_{2M_{\odot}}$ is very poor (0.04). As explained in HTZCS, at high densities the isoscalar vector field $\omega$ grows linearly with density, while the equation of motion for the isovector vector field $\rho$ has a trivial solution for a $\rho$ field growing inversely proportional to the density. For a 2M$_{\odot}$ neutron star, these two effects lead to a cancellation of the density dependence of the nuclear symmetry energy, i.e. the slope $L_{sym}$ around saturation density. So, we expect very little correlation of $L_{sym}$ with observables for a 2M$_{\odot}$ NS. 
   \end{itemize}

\section{Results: Full Analysis}
\label{sec:full}

In HTZCS, the isoscalar saturation parameters were kept fixed and only the isovector parameters varied within their uncertainties. It is however unclear whether simultaneous variation of all the parameters would affect the above conclusions. This is crucial, as a study of sensitivity to each individual empirical parameter does not take into account any underlying correlations among the parameters themselves, or with the observables.
\\

In this section, we will vary both isoscalar and isovector nuclear empirical parameters simultaneously within their uncertainty ranges, motivated by state-of-the-art nuclear experimental data. We will then impose additional constraint filters from heavy-ion collision experiments to further restrict the parameter space. Finally we will investigate possible correlations among the empirical nuclear parameters and the astrophysical observables. 
\\

\subsection{Variation of all nuclear parameters}
Having established our scheme and verified the results of HZTCS, now we extend our analysis to all the nuclear parameters and also for larger number of parameter sets. The ranges of variation for the nuclear parameters are taken to be compatible with estimations coming from the analysis of a variety of nuclear data from terrestrial experiments, astrophysical observations and theoretical calculations.
The compressibility modulus $K_{sat}$ is taken to be in the range 200-300 MeV, as obtained from current experimental data on isoscalar giant monopole and dipole resonances (compression
modes) in nuclei \cite{Shlomo,Piekarewicz}. For the energy per particle sat saturation $E_{sat}$, we adapt values -16$\pm$0.2 MeV  from binding energy of atomic nuclei~\cite{Atkinson,Kejzlar}. The symmetry energy $E_{sym}$ is varied between 28-34 MeV, the slope of symmetry energy $L_{sym}$ in the range 40-70 MeV, while the nucleon effective mass m*/m is considered to vary between 0.55-0.75 as in HTZCS. The ranges of all the nuclear parameters considered in the full analysis are listed in Table~\ref{tab:parafull}.
\\

\begin{table*} [htbp]
\caption{Range of empirical parameters considered for the full analysis in Sec.~\ref{sec:full}.}
\label{tab:parafull}       
\begin{center}
\begin{tabular}{llllll}
\hline\noalign{\smallskip}
   $n_0$ & $E_{sat}$ & $K_{sat}$ & $E_{sym}$ & $L_{sym}$ & $m^*/m$ \\
 ($fm^{-3}$) & (MeV) & (MeV) & (MeV) & (MeV) & {}\\
\noalign{\smallskip}\hline\noalign{\smallskip}
{0.14 - 0.17} & { -16$\pm$0.2} & {200 - 300} & {28 - 34} & {40 - 70} & {0.55 - 0.75} \\
\noalign{\smallskip}\hline
\end{tabular}
\end{center}
\vspace*{1cm}  
\end{table*}

\begin{figure}[h]
  \begin{center}
  \resizebox{0.55\textwidth}{!}{%
      \includegraphics{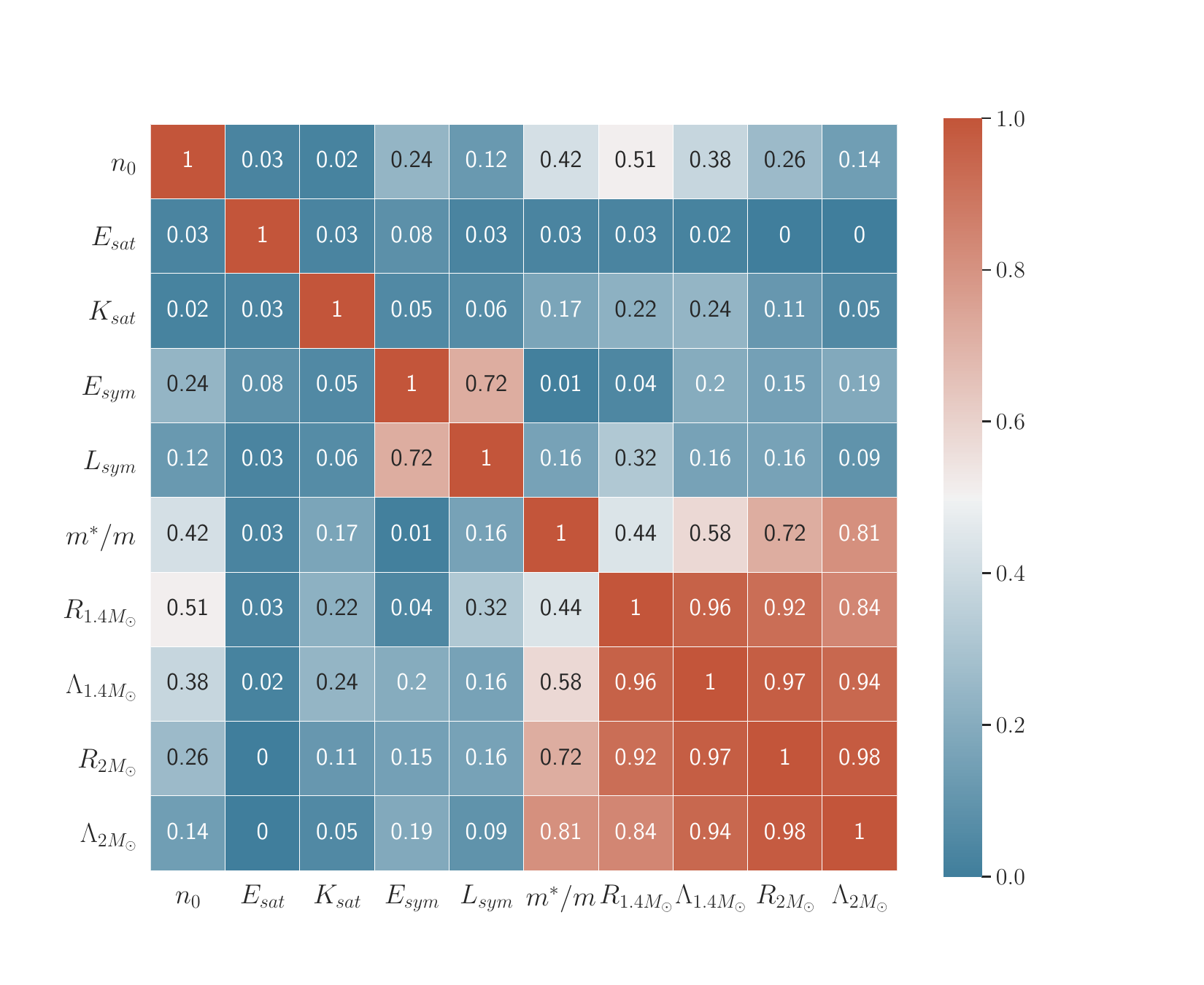}
      }
    \caption{Posterior correlation matrix for variation of all nuclear empirical parameters and NS observables, after application of the $\chi$EFT and NS observations filter (maximum mass of  PSR J0740+6620~\cite{fonseca2021refined} and tidal deformability $\Lambda$ from  GW170817~\cite{Abbott2019})}
    \label{fig:corrfigall}
  \end{center}
\end{figure}

Following the same cut-off filter scheme described in Sec.~\ref{sec:bayesian}, we now obtain posteriors of 1000 points for the nuclear parameters and NS observables, for the filters from $\chi$EFT and NS observations. 
In Fig.~\ref{fig:corrfigall}, we display the correlation matrix of the following quantities:
nuclear empirical parameters ($n_0$, $E_{sat}$, $K_{sat}$, $E_{sym}$, $L_{sym}$), the
effective mass $m^*/m$ and the NS observables ($R_{1.4M_{\odot}}$, $\Lambda_{1.4M_{\odot}}$, $R_{2M_{\odot}}$, $\Lambda_{2M_{\odot}}$). It was found that the correlation values between the variables do not significantly change with increasing the number of posteriors points. Some of the main observations from the correlation matrix are listed below: 
\begin{itemize}
    \item 
    $n_0$ and $m^*/m$ now show a moderately good correlation (0.42).
    \item $L_{sym}$ and $E_{sym}$ display a strong correlation (0.72), higher than our observation in the previous section
    \item  $n_0$ has a weak correlation with the NS observables. The correlation is noticeable (0.51) for 1.4M$_{\odot}$ NS but is negligible for 2~M$_{\odot}$.
    \item The correlation of $m^*/m$ with 1.4~M$_{\odot}$ NS observables is moderate ($\approx$ 0.5), but is higher for 2~M$_{\odot}$ NS.
    \item The correlation between $L_{sym}$ and radius of 1.4M$_{\odot}$ NS is also lower (0.32) than our previous findings.
    \item All the NS observables (radius and dimensionless tidal deformability for 1.4~M$_{\odot}$ and 2~M$_{\odot}$ NS), as expected, show a strong correlation with each other (according to Eq.~\ref{eq:tidal}).
\end{itemize}


\subsection{Correlations with statistical weighting}
\label{sec:weight}

\begin{figure}[h]
  \begin{center}
  \resizebox{0.55\textwidth}{!}{%
      \includegraphics{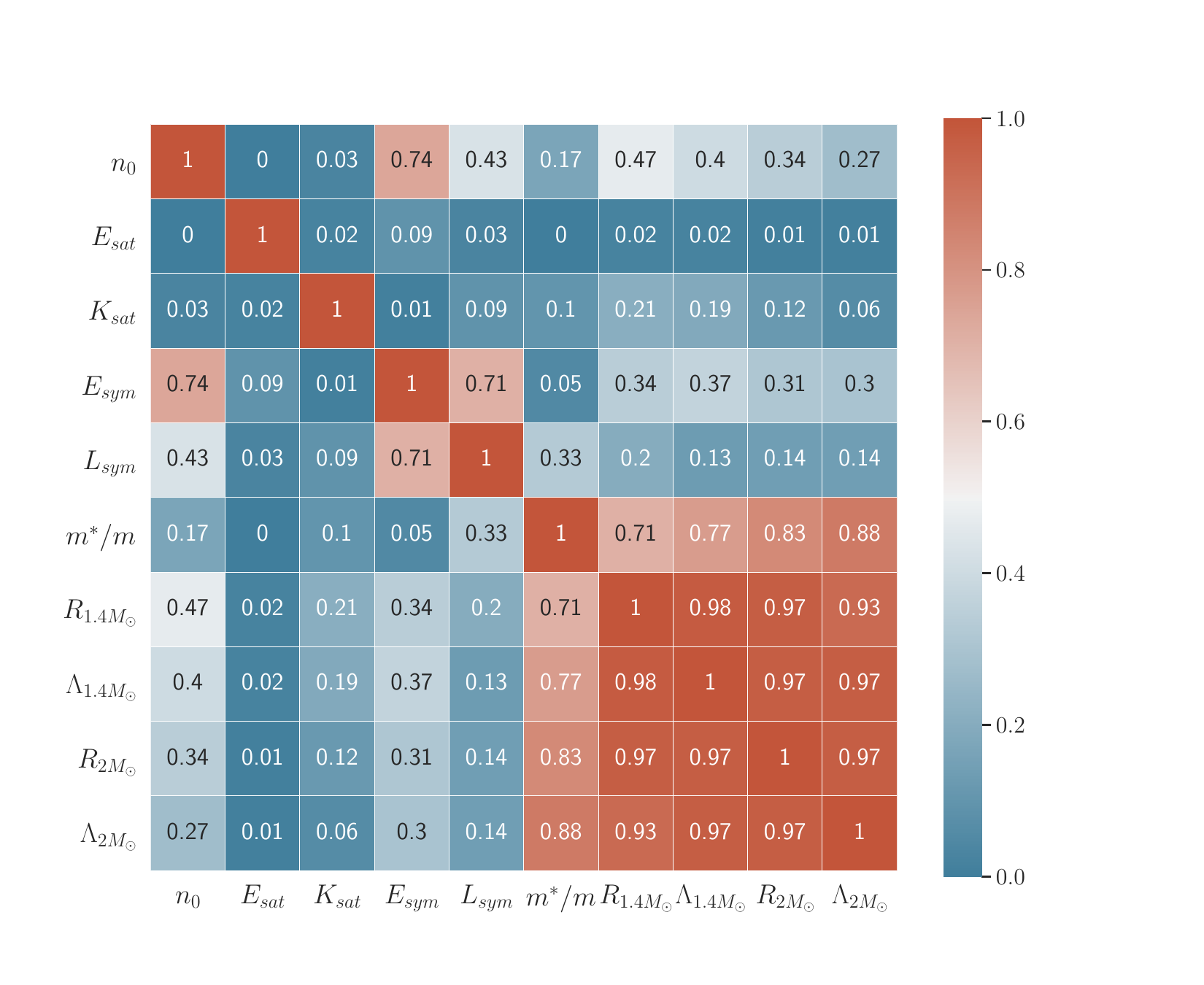}
      }
\caption{Same as Fig.~\ref{fig:corrfigall}, after inclusion of statistical weighting}
    \label{fig:corrfigall_weight}
  \end{center}
\end{figure}

As discussed in Sec.~\ref{sec:bayesian}, one must check whether the inclusion of statistical weighting in the applied scheme with the hard cut-off limit considered significantly affects the conclusions of this study. In order to do that, we recalculate the posterior correlation matrix for variation of all nuclear empirical parameters and NS observables, after application of the $\chi$EFT and NS observations filters as in Fig.~\ref{fig:corrfigall}, by including a proper calculation of the likelihood of the data for the given model. The results are shown in Fig.~\ref{fig:corrfigall_weight}.

On comparison with Fig.~\ref{fig:corrfigall}, the following conclusions can be drawn:
\begin{itemize}
 \item The correlation between $E_{sym}$ and $L_{sym}$ remains unaltered (0.71).
    \item The correlations of $m^*/m$ with observables $R_{1.4 M_{\odot}}$ and $\Lambda_{1.4 M_{\odot}}$ increase to 0.71 and 0.77 respectively, and with $R_{2 M_{\odot}}$ and $\Lambda_{2 M_{\odot}}$ remain strong (0.83 and 0.88 respectively).
    \item The correlation between $n_0$ and $m^*/m$ has decreased to 0.17.
    \item A strong correlation now appears between $n_0$ and $E_{sym}$ (0.74) while that with $L_{sym}$ increases to 0.43.
\end{itemize}

In summary, other than the expected correlations among NS observables and between symmetry energy and its slope, we found strong correlations of the effective nucleon mass with the NS observables and between saturation density and the symmetry energy.

\subsection{Additional constraints from heavy-ion collision experiments}
\label{sec:hic}

In the previous section, we imposed constraints on the nuclear parameters at very low-density from $\chi$EFT data and at very high density from multi-messenger observations of neutron stars. Now, in this section we will apply the results from several other heavy-ion collision experiments, which impose important constraints on the nuclear parameters in the density region of $n_b/n_0 \sim 1-3$. However, it is well known that the above constraints are model-dependent, and therefore the implications of the results must be taken with caution. 

\subsubsection{KaoS Experiment}
\label{sec:KaoS}
As discussed in Sec.~\ref{sec:filters}, results from the KaoS collaboration directly test the EoS for densities $\sim 2-3 n_0$. 
For Skyrme type models this restriction implies compression moduli of $K_{sat} \leq$ 200 MeV~\cite{GORIELY2005425,CHABANAT1998}. The constraint given by the KaoS data is accomplished by only allowing nucleon potentials which are more attractive than the Skyrme parametrization within the considered density regime. Therefore for RMF models, one should derive the nucleon potential and compare it with the Skyrme parametrization with $K_{sat} \leq$ 200 MeV. A more attractive potential at high densities will allow matter to be compressed more for the same incident energy, enhancing multiple scattering processes in subthreshold kaon production~\cite{Sagert}.
\\

\begin{figure}[h]
  \begin{center}
  \resizebox{0.5\textwidth}{!}{%
{\includegraphics{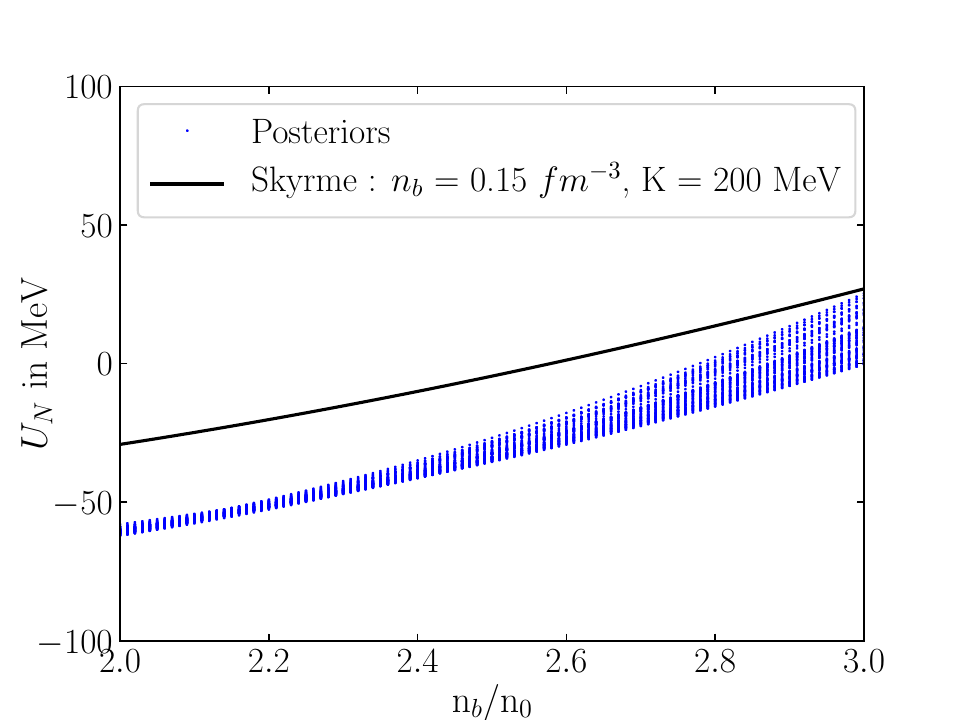}}
}
      \caption{Nucleon potential of the posterior ANM EoSs allowed by KaoS Experiment~\cite{Hartnack}}
    \label{fig:UNposterior} 
  \end{center}
\end{figure}

In this analysis the non-relativistic Skyrme model is used to produce the EoS for $K_{sat}$=200 MeV and $n_0 = 0.15 fm^{-3}$. To impose the KaoS constraint, we calculate the nucleon potentials for the priors (saturation parameters randomly picked from the parameter space in Table~\ref{tab:emppara} as described in Sec.~\ref{sec:prior}) and allow only those for which the nucleon potential is less than that for the Skyrme EoS. The results are shown in Fig.~\ref{fig:UNposterior}. 
Around 40\% of our prior sets pass through this filter.

\subsubsection{FOPI Constraint}
\label{sec:fopi}
\begin{figure}[h]
  \begin{center}
  \resizebox{0.5\textwidth}{!}{%
      \includegraphics{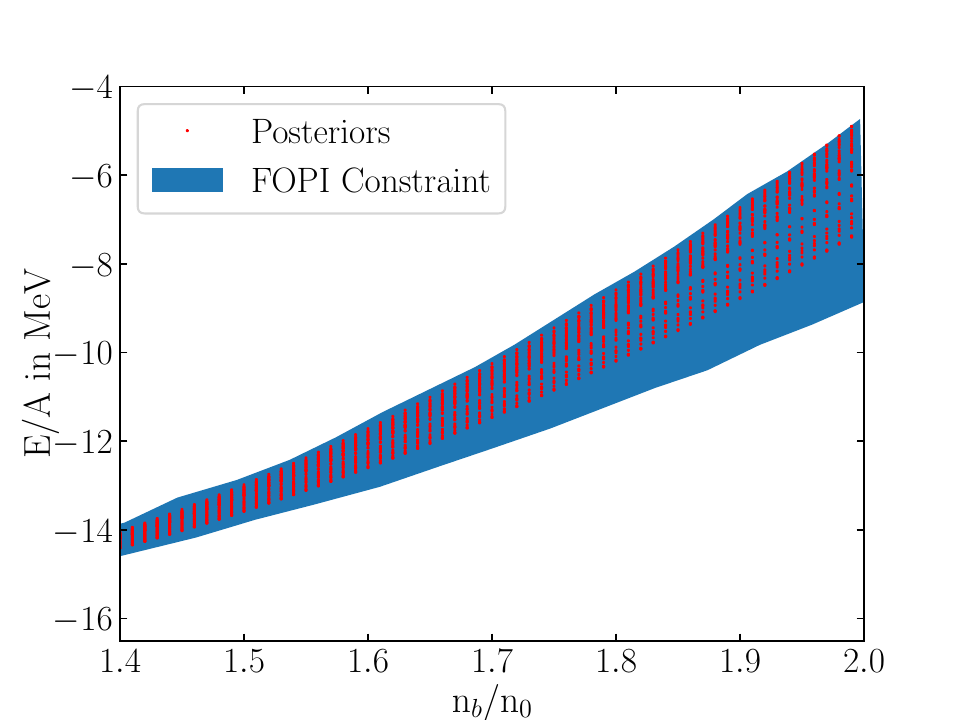}
      }
      \caption{Binding energy of the posterior SNM EoSs allowed by the FOPI data~\cite{FOPI}}
    \label{fig:FOPIposterior} 
  \end{center}
\end{figure}

In Sec.~\ref{sec:filters}, we discussed the fact that using the FOPI data, one can obtain a constraint for the binding energy of SNM in the density region of 1.4-2 $n_0$. To impose this constraint, we calculate the binding energy for SNM for the input parameters randomly picked from the parameter space as in Table~\ref{tab:parafull} and allow only those for which the binding energy lies inside the band allowed by the FOPI data in this density range. This is depicted in Fig.~\ref{fig:FOPIposterior}. Nearly 20\% of our prior sets pass through the FOPI filter. \\

\subsubsection{ASY-EOS constraint}
\label{sec:asy-eos}
\begin{figure}[h]
  \begin{center}
    \resizebox{0.5\textwidth}{!}{%
      \includegraphics{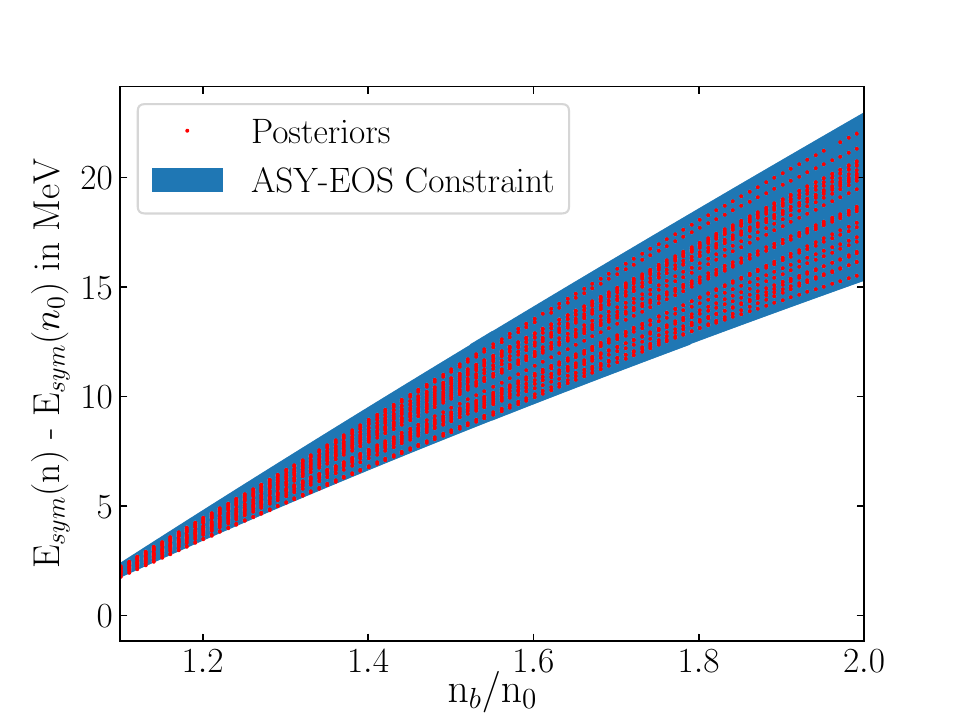}
      }
      \caption{Symmetry energy of the posteriors EoSs with ASY-EOS band~\cite{ASY_EOS}}
    \label{fig:ASYEOSposterior} 
  \end{center}
\end{figure}

It was discussed in Sec.~\ref{sec:filters} that information about the symmetry energy for ANM at supra-saturation density can be obtained from the recent ASY-EOS data. We computed the symmetry energy for ANM EoS using Eq.~\ref{eqn:symmenergy} below~\cite{GlendenningBook}:

\begin{equation}\label{eqn:symmenergy}
    E_{sym} = \frac{k_F^2}{6\sqrt{k_F^2+m^{*2}}} +\left(\frac{g_{\rho}^2}{m_{\rho}^2 + 2\Lambda_{\omega}g_{\rho}^2g_{\omega}^2\omega_0^2}\right) \frac{k_F^3}{12\pi^2}~,
\end{equation}
where $k_F$ is the Fermi momentum of the nucleons. 
Since the value of the symmetry energy at saturation in the interval between 30 MeV and 32 MeV seem to be favoured by a majority of terrestrial experiments and astrophysical observations ~\cite{Bao,Lattimer2014}, we used the constraints on $E_{sym}$ for the choice of parametrization at $E_{sym}(n_0)$ = 31 MeV as in~\cite{ASY_EOS}. To impose the constraint filter in the Bayesian scheme, we calculate the symmetry energy of ANM EoS using Eq.~\ref{eqn:symmenergy} for the input parameters randomly picked from the parameter space as in Table~\ref{tab:parafull} and allow only those for which the symmetry energy lies inside the band allowed by the ASY-EOS data in the range of $\sim 1.1 - 2.0 n_0$ (Fig.~\ref{fig:ASYEOSposterior}). We find that around 40\% of our prior sets pass through this filter.

\subsection{Correlation matrix with all constraints: $\chi$EFT + GW observables + HIC (KaoS, FOPI \& ASY-EOS)}
\label{sec:corrHIC_noweights}

In Fig.~\ref{fig:corrfigall}, we displayed the correlation matrix for all the nuclear parameters and the NS observables for the posteriors obtained after passing through the filters from $\chi$EFT and NS observations. Now we will add three additional HIC constraints from KaoS, FOPI and ASY-EoS experiments discussed in Sec.~\ref{sec:filters}. On applying the constraint from KaoS  experiment in addition to $\chi$EFT and NS observables, the correlation of $n_0$ with the NS observables increases while that with $m^*/m$ decreases. The correlation between $K_{sat}$ and $m^*/m$ also increases (not displayed here).
\\

\begin{figure}[h]
  \begin{center}
    \resizebox{0.55\textwidth}{!}{%
      \includegraphics{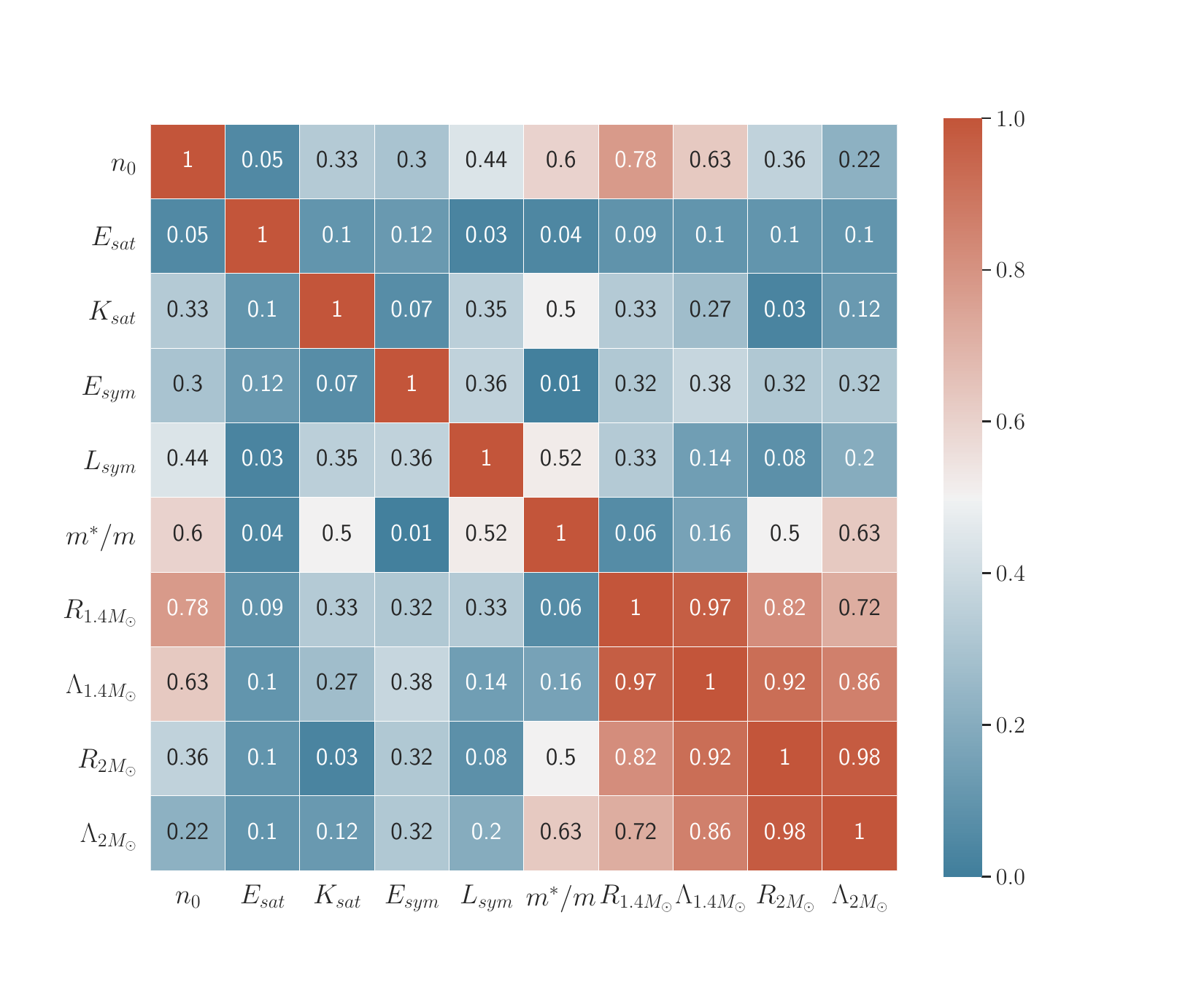}
      }
    \caption{Posterior correlation matrix for all the nuclear empirical parameters and NS observables, after application of the $\chi$EFT,HIC experiments (KaOS, FOPI, ASY-EOS)  and NS observations filters (maximum mass of  PSR J0740+6620~\cite{fonseca2021refined} and tidal deformability $\Lambda$ from  GW170817~\cite{Abbott2019})}
    \label{fig:corrfigall5}
  \end{center}
\end{figure}

After adding all the HIC filters, we reproduce the correlation plot given in Fig.~\ref{fig:corrfigall5}. There are a few differences between the plots before (Fig.~\ref{fig:corrfigall}) and after applying the constraints from KaoS, FOPI and ASY-EOS experiments (Fig.~\ref{fig:corrfigall5}):
\begin{enumerate}
    \item The correlation of $n_0$ with other nuclear parameters except $E_{sat}$ has increased.
   \item The correlation of $L_{sym}$ with $E_{sym}$ has decreased (almost by half to 0.36) and now $L_{sym}$ has a significant correlation (0.52) with $m^*/m$.
   \item The correlation among NS observables is not significantly altered due to the HIC filters.
   \item $m^*/m$ now has negligible correlation with the observables for 1.4~M$_{\odot}$ stars but correlation with $n_0$ (0.78) has increased from previous case.
\end{enumerate}

\subsection{Correlations with all ($\chi$EFT+astro+HIC) constraints including weighting}
\label{sec:weight_hic}

In Sec.~\ref{sec:weight}, we studied the correlations applying filters from $\chi$EFT and NS astrophysical observables, including statistical weighting in the applied scheme. In the previous section, we included the constraints from HIC experiments, and studied their effects on the correlations (without statistical weighting).
As discussed in Sec.~\ref{sec:bayesian}, inclusion of weights for HIC may introduce a false confidence in the results, as HIC experiments are known to be model dependent and have large uncertainties. In this section, we proceed to calculate the correlations including all filters ($\chi$EFT+astro+HIC) and with statistical weighting, with a word of caution. 
\\

\begin{figure}[h]
  \begin{center}
    \resizebox{0.55\textwidth}{!}{%
      \includegraphics{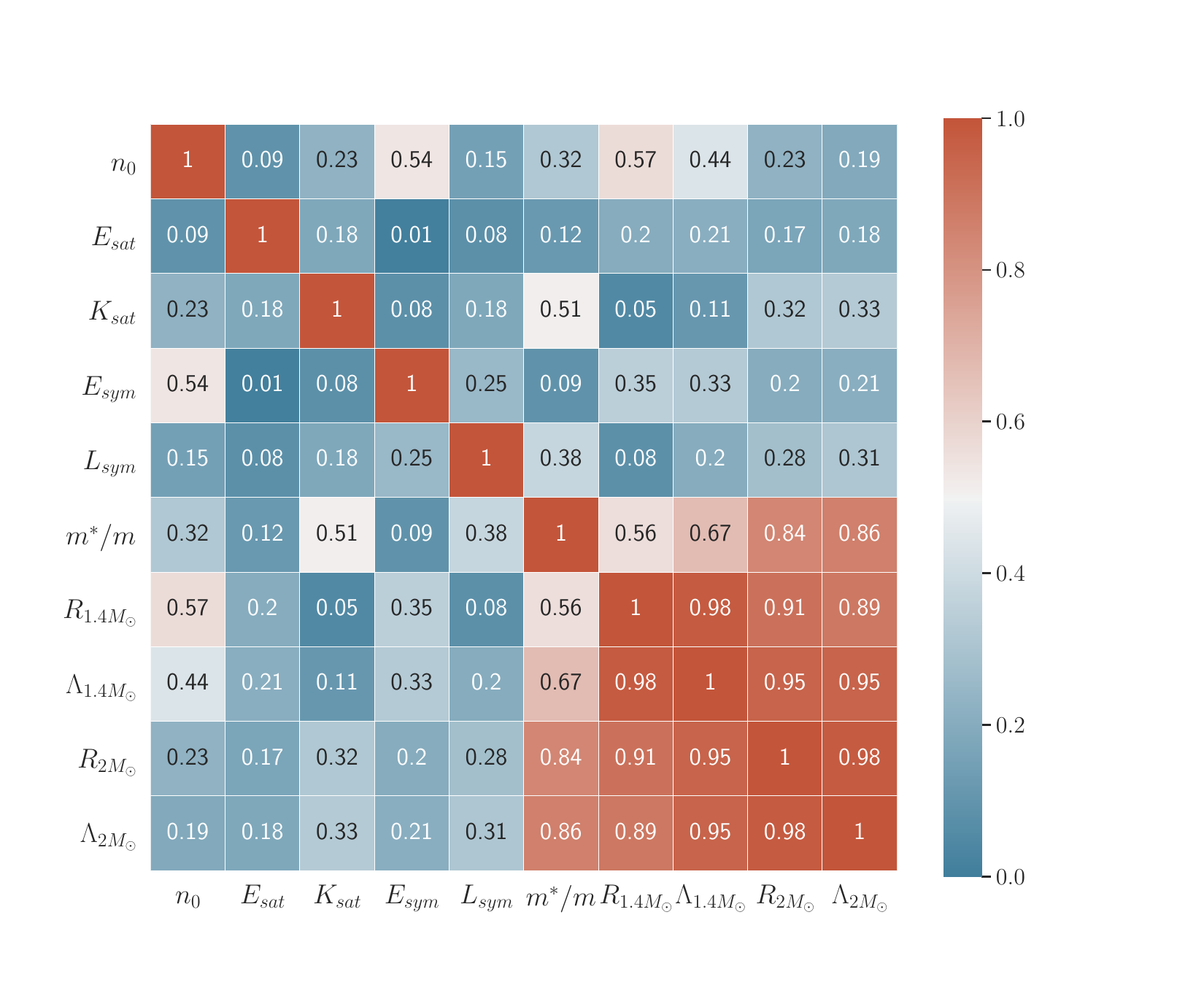}
      }
    \caption{Same as in Fig.~\ref{fig:corrfigall5} after inclusion of statistical weighting}
    \label{fig:corrfigall5_wt}
  \end{center}
\end{figure}

In Fig.~\ref{fig:corrfigall5_wt}, we display the resulting correlations. Upon comparison with Fig.~\ref{fig:corrfigall_weight}, we can draw the following conclusions:
\begin{enumerate}
    \item The correlation of $n_0$-$E_{sym}$ (0.54) still persists, while the  $n_0$-$L_{sym}$ correlation weakens (0.15)
    \item The correlation of $L_{sym}$ and $E_{sym}$ changes significantly on application of HIC filters, from 0.71 to 0.25
    \item As in Fig.\ref{fig:corrfigall_weight}, there is a weak correlation of $n_0$ with NS observables  
    \item The correlation of $K_{sat}$ with $m^*/m$ has increased from 0.1 to 0.51
    \item The $(L_{sym},m^*/m )$ correlation is not significantly affected
\end{enumerate}

\section{Discussions}
\label{sec:discussions}

\subsection{Summary of the results}
\label{sec:summary}
In this work we applied the current best-available information from multiple channels, namely nuclear theory and experiments, heavy-ion collision experiments as well as multi-messenger (multi-wavelength electromagnetic as well as GW) astrophysical observations to constrain the nuclear EoS at different densities. The phenomenological RMF model applied here satisfies state-of-the-art nuclear matter saturation properties, low-density constraints from neutron matter from $\chi$EFT ab-initio approach, heavy-ion collision experimental results at intermediate densities and multi-messenger astrophysical observational data at high densities. We applied the constrained EoS parameter space to study correlations among nuclear empirical parameters and NS observables by applying a cut-off filter scheme. 
\\

In order to test the approach, we reproduced the results of the recent investigation by Hornick et al.~\cite{Hornick} (HTZCS). We simultaneously varied the isovector nuclear empirical parameters within the allowed parameter space of uncertainties, to generate random EoSs. Then we imposed filters from physical constraints such as $\chi$EFT data and recent NS astrophysical observations of mass, radius and tidal deformability to restrict the choice of parameters. We used the filtered parameter sets to study correlations among the parameters themselves and also with the NS observables. We found strong correlations between $E_{sym}$ and $L_{sym}$, in agreement with previous literature. We confirmed the findings of HTZCS that the correlation of $R_{1.4 M_{\odot}}$ with $m^*/m$ is stronger than that of $L_{sym}$. Our results also agreed with HTZCS that the effective nucleon mass is highly correlated with the radius and tidal deformability. The expected strong correlation between tidal deformability and radius was also seen for both 1.4 and 2 $M_{\odot}$ stars.
\\

The scheme was then extended to a full analysis, considering variation of both isoscalar and isovector saturation parameters and also for a larger sample of data in order to improve the statistics and to test the robustness of the results. 
We observed the usual strong correlation of $L_{sym}$ with $E_{sym}$, and a weaker correlation of $L_{sym}$ with $R_{1.4 M_{\odot}}$. We also recovered the expected correlations among the NS radii and tidal deformability. In addition, we also found weak correlations of $n_0$ with $m^*/m$, and with the NS observables. 
\\

\begin{figure}[htbp]
  \begin{center}
  \resizebox{0.5\textwidth}{!}{%
      \includegraphics{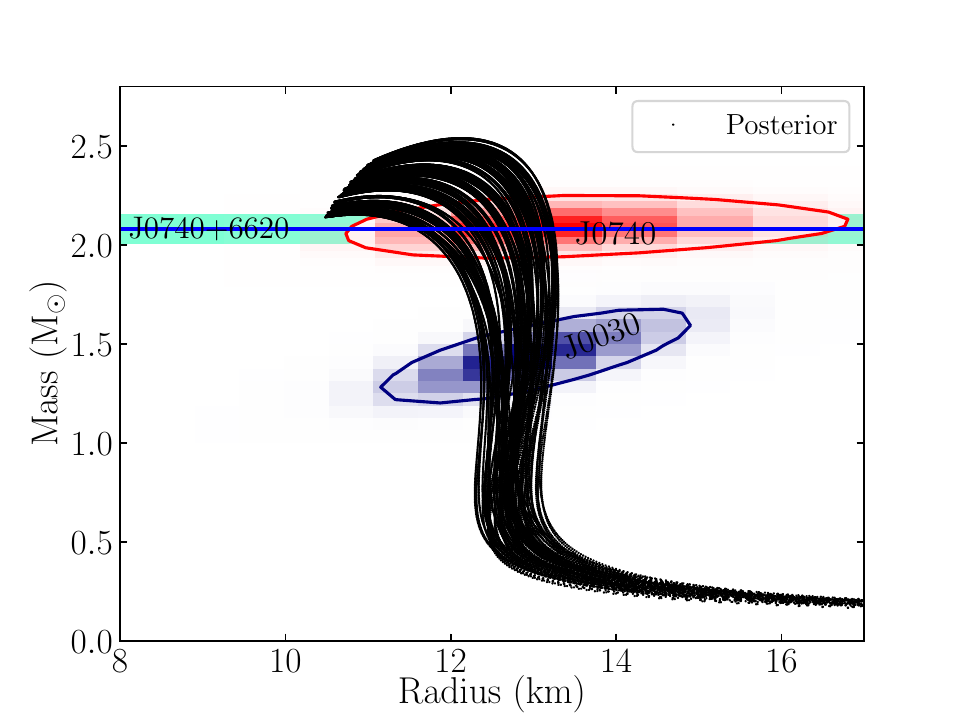}
      }
    \caption{Mass-radius relations for posterior ANM EoSs. The elliptical red and deep blue regions denoted by lines correspond to 1-$\sigma$ limits of the mass and radius measurement of the pulsar PSR J0740+6620~\cite{NICER0740_Miller} and PSR J0030+0451~\cite{NICER0030_Miller} respectively. The light green band indicates the uncertainty in the measurement of maximum mass of PSR J0740+6620~\cite{fonseca2021refined}.}
    \label{fig:mr_nicer}
  \end{center}
\end{figure}

NICER data provide constraints on the mass and radius of a pulsar. The Bayesian parameter estimation of the mass and equatorial radius of the millisecond pulsar PSR J0030+0451 ~\cite{NICER0030_Miller} yield $M=1.44^{+0.15}_{-0.14}$ M$_{\odot}$ and $R_e=13.02^{+1.24}_{-1.06}$ km, consistent with the independent analysis by \cite{NICER0030_Riley}. More recently for PSR J0740+6620, the equatorial circumferential radius obtained was $13.7^{+2.6}_{-1.5}$ km~\cite{NICER0740_Miller} (68\%). In Fig~\ref{fig:mr_nicer}, we superimpose the filtered M-R curves with the mass and radius measurements from the recent NICER data, for pulsar PSR J0030+0451  \cite{NICER0030_Miller,NICER0030_Riley} and PSR J0740+6620 \cite{NICER0740_Miller}. 
We verified that our results are consistent with the latest analyses from NICER. All our filtered points pass through the 1-$\sigma$ limit of the M-R region of the NICER observations for pulsar PSR J0030+0451  \cite{NICER0030_Miller} and PSR J0740+6620 \cite{NICER0740_Miller}. However, another independent radius measurement for J0740+6620 was performed by Riley~et~al.(2021)~\cite{NICER0740_Riley}, which differs from the Miller~et~al.(2021) results~\cite{NICER0740_Miller} by $\approx$ 0.8 km at the $1\sigma$ credibility contour. Some of our M-R posterior curves do not agree with the analysis by Riley et al.~\cite{NICER0740_Riley} within 1-$\sigma$ but all posteriors agree within 2-$\sigma$ (not shown here).
\\

\begin{figure}[htbp]
  \begin{center}
  \resizebox{0.5\textwidth}{!}{%
      \includegraphics{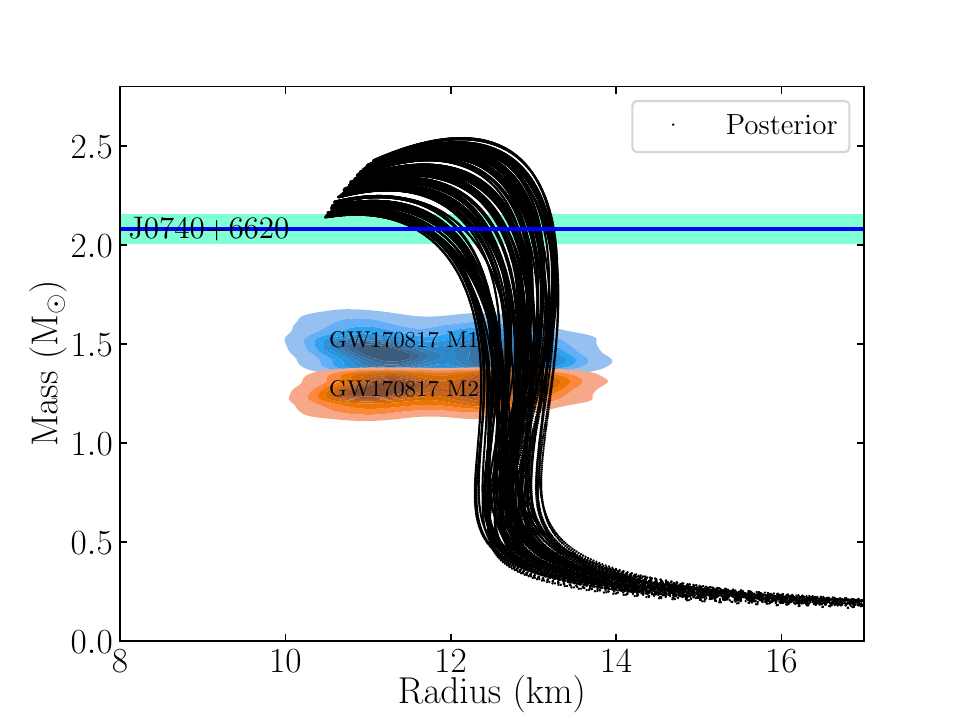}
      }
    \caption{Mass-radius relations for posterior ANM EoSs. The light blue and orange regions above and below the mass value of
M=1.365 M$_{\odot}$ correspond to the mass-radius estimates for the two compact stars of the merger in GW170817~\cite{Abbott2018,LSCData_GW170817}. The light green band indicates the uncertainty in the measurement of maximum mass of PSR J0740+6620~\cite{fonseca2021refined}.}
    \label{fig:mr_GW170817}
  \end{center}
\end{figure}

In Fig~\ref{fig:mr_GW170817}, we superimpose the M-R curves with the posterior mass-radius relations of the two neutron stars from GW170817 observations ~\cite{Abbott2018,LSCData_GW170817} obtained by using a low-spin prior and parametrized-EoS with constraint of maximum mass of 1.97$M_{\odot}$ for comparison. We find that that all our posteriors points are within the 90\% confidence interval. This study does not use posteriors from previous LIGO-Virgo Collaboration publications, and any correlations that may be implicit in such posteriors have not been considered. The correlations studied here are calculated from the confidence bounds used in this study.
\\

We then investigated the effect of imposing additional constraints from heavy-ion data and investigated their impact on the EoS at intermediate densities ($\sim 1-3 n_0$). Applying only the KaoS constraint increases the correlation of $n_0$ with the NS observables while decreasing the same for $m^*/m$. It also increases the correlation of incompressibility $K_{sat}$ with $m^*/m$. We further studied the combined effect of imposing all the HIC filters (KaoS, FOPI and ASY-EOS experiments) in addition to the $\chi$EFT and NS observational constraints.
The two considerable changes in correlation with application of HIC filters are decreasing of $E_{sym}$ and $L_{sym}$ correlation and increase in the correlation of $m^*/m$ and $K_{sat}$. The first one can be attributed to either or both of FOPI and ASY-EOS.
\\

Further, we verified whether the inclusion of statistical weighting for the filters affect the conclusions significantly. Other than the expected correlations among NS  observables  and  between  symmetry  energy  and  its slope, we found strong correlations of the effective nucleon mass with the NS observables and between saturation density and the symmetry energy. The latter was found to weaken on inclusion of heavy-ion filters with corresponding statistical weighting, but this data is known to be model-dependent. In summary, the most important nuclear parameters to consider for astrophysical data are the effective nucleon mass $m^*/m$ and the nuclear saturation density $n_0$.

\subsection{Comparison with previous studies}
\label{sec:comparison}

NS radii are known to be sensitive to the symmetry energy at around two times the saturation density~\cite{Lattimer2000,Lattimer2001}. It was first discussed by Fattoyev et al.~\cite{Fattoyev2014} that the tidal deformability of low mass neutron stars is  sensitive to $L_{sym}$, while the tidal deformability of massive neutron stars is sensitive to the high density behavior of symmetry energy. A possible relation between $L_{sym}-R_{1.4 M_{\odot}}-\Lambda_{1.4 M_{\odot}}$ was searched by Zhang et al.~\cite{Zhang2018,Zhang2020}. Hornick et al. \cite{Hornick} found $R_{1.4 M_{\odot}}$ is almost independent of $L_{sym}$ on applying  constraints from $\chi$EFT. Using several non-relativistic and relativistic RMF and Skyrme Hartree-Fock (SHF) models, Malik et al.~\cite{Malik2018} found that the tidal deformability is weakly or only moderately correlated with the individual nuclear matter parameters of the EoS. They also found stronger correlations of $\Lambda$ for specific choices of the linear combinations of the isoscalar and isovector EoS parameters.
\\

The strong correlation between $L_{sym}$ and $E_{sym}$ as well as $L_{sym}$ and $R_{1.4 M_{\odot}}$ were reported in many works~\cite{Guven2020} combining different constraints such as astrophysical data from LIGO/Virgo or NICER, $\chi$EFT theory and neutron skin thickness from PREX II results~\cite{Fattoyev2018}. A mild tension was reported between nuclear empirical parametrization with those from astrophysical and $\chi$EFT results~\cite{essick2021astrophysical}. Other works which have employed a hybrid parametrization (nuclear+piecewise polytrope) have not found such apparent tension ~\cite{biswas2021prex,Reed2021} and found a strong correlation between $R_{1.4M_{\odot}}$ and $L_{sym}$ based on PREX II results.
\\

Recently Huth~et~al.~\cite{huth2021} combined data from 
astrophysical multi-messenger observations of neutron stars and from heavy-ion collisions (FOPI~\cite{FOPI} and ASY-EOS~\cite{ASY_EOS} experiments as well as EoS constraint for symmetric nuclear matter \cite{Danielewicz}) with microscopic $\chi EFT$ calculations and used a Bayesian inference technique to analyse the nuclear EoS and NS properties. While $\chi EFT$ was used to constrain the EoSs at densities below 1.5n$_0$, an extrapolation using the speed-of-sound in NS matter was used to extend the EoS at higher densities, with the additional criterion to support at least 1.9M$_{\odot}$ NS mass. The mass measurements
of the PSR J0348+404232 and PSR J1614-22303 were used to obtain lower bound and the NS in GW170817 to obtain an upper bound for the NS maximum mass. Other multi-messenger information such as from NICER and XMM Newton missions, as well as GW data and kilonova AT2017gfo9 associated with the
GW signal were also used in the Bayesian framework. The final EOS constraints were obtained through the combination
of both the HIC information and astrophysical multi-messenger observations. The study concluded that the constraints from HIC experiments is in excellent agreement with NICER observations. In comparison with our work, Huth~et~al.~\cite{huth2021} start with a soft EOS from chiral EFT up to 1.5 $n_0$ and the EOS becomes stiffer due to the large $L_{sym}$ values of the ASY-EOS predictions, and the final results only slightly depend on the EOS of SNM. In contrast for the RMF model used in our work, RMF models are more isospin symmetric at high density so that they are less sensitive to parameters of PNM. 
\\

\subsection{Implications and outlook}
\label{sec:implications}

We must recall here the difference between the astrophysical versus the heavy-ion constraints applied here and their impact on the parameters. The constraints from heavy-ion data are model-dependent and therefore we treat them separately from astrophysical constraints and discuss the implications of the results with a word of caution. We also note here that the EoS beyond 2-3 $n_0$ can be associated with the emergence of new degrees of freedom, such as the appearance of hyperons or quarks which we have not considered in this study. With the upcoming CBM experiment at FAIR~\cite{CMBbook} the EoS in this density regime can be probed further in the near future.
\\

In conclusion, nuclear empirical parameters control important properties of nuclear matter from which accurate predictions for dense matter EoS can be made based on our current knowledge, improved by up-to-date constraints from experimental and astrophysical data as well as ab-initio approaches.
Numerical relativity simulations of astrophysical scenarios such as supernovae or binary NS mergers require cold EoSs in beta-equilibrium for the initial data. In order to probe the effect of the microphysics (e.g. new degrees of freedom) on NS astrophysical observables, one requires a large variability of EoSs covering the entire parameter space. The knowledge of the constrained EoS parameter space resulting from simultaneous application of constraints, coming from different physics at different density regimes, can help to make an informed choice of the parameters in the simulation codes. 
\\

In the present work, we have pointed out the most important empirical parameters which are mainly responsible for the uncertainty in the NS observables based on the nucleonic EoS. According to the results of our investigation, the essential parameters for RMF models to vary are the saturation nuclear density $n_0$ and effective nucleon mass $m^*/m$. This was also concluded in a recent investigation of the effect of dominant RMF parameters on $f$-mode oscillations in neutron stars~\cite{Bikram2021}. This conclusion results from the fact that $n_0$ showed a moderate correlation with $R_{1.4 M_{\odot}}$, while $m^*/m$ showed a strong correlation with NS observables like radius and tidal deformability.  However on imposing additional constraints from heavy-ion data, the correlation of $n_0$ with $R_{1.4 M_{\odot}}$ increased while that of $m^*/m$ decreased, although it is noted that HIC constraints are less robust and model-dependent. These results should therefore be of interest to the community performing astrophysical and numerical relativity simulations.

\section*{Acknowledgements}
\label{sec:ack}

DC is grateful to the Alexander von Humboldt Foundation for the Renewed Research Stay grant (Nov 2019 - Jan 2020) and the warm hospitality of the Institut f\"ur Theoretische Physik (ITP), Goethe Universit\"at in Frankfurt am Main, Germany where majority of this research project was done.
JSB acknowledges support by the Deutsche Forschungsgemeinschaft (DFG, German Research Foundation) through the CRC-TR 211 'Strong-interaction matter under extreme conditions'– project number 315477589 – TRR 211.
The authors also thank Christian Drischler for providing the $\chi$EFT  data used in this work. DC deeply acknowledges calculations by Jan-Erik Christian and Andreas Zacchi at ITP for comparison of the results. SG is thankful to Bikram Pradhan and Bhaskar Biswas for the useful discussion sessions they had during the project.

\bibliographystyle{apsrev4-1}
\bibliography{AvH.bib}

\end{document}